\documentstyle[twocolumn,psfigure,rotating]{article}
\newcommand{\hrc}{{\tt HRC}}
\newcommand{\h}{{\sc HERMES}}

\begin{document}

\onecolumn
\setcounter{page}{1}

\LARGE
\centerline{\bf The HERMES Spectrometer}
\normalsize

\bigskip

\begin{center}

K.~Ackerstaff$^{5}$, 
A.~Airapetian$^{30}$, 
N.~Akopov$^{30}$, 
M.~Amarian$^{30}$, 
V.~Andreev$^{24}$, 
E.C.~Aschenauer$^{11,21}$, \\
R.~Avakian$^{30}$, 
H.~Avakian$^{9}$, 
A.~Avetissian$^{30}$, 
B.~Bains$^{13}$
% ...

\bigskip

% ...
$^5$DESY, Deutsches Elektronen Synchrotron, 22603 Hamburg, Germany\\
% ...
$^{9}$Istituto Nazionale di Fisica Nucleare, Laboratori Nazionali di Frascati,
I-00044 Frascati, Italy\\
% ...
$^{11}$Department of Subatomic and Radiation Physics, University of Gent, 
9000 Gent, Belgium\\
% ...
$^{13}$Department of Physics, University of Illinois, Urbana, 
IL 61801, USA\\
% ...
$^{21}$Nationaal Instituut voor Kernfysica en Hoge-Energiefysica (NIKHEF), 
1009 DB Amsterdam, Netherlands\\
% ...
$^{24}$Petersburg Nuclear Physics Institute, St.Petersburg, 188350 Russia\\
% ...
$^{30}$Yerevan Physics Institute, 375036, Yerevan, Armenia

\bigskip

\begin{abstract}
The HERMES experiment is collecting data on inclusive and
semi-inclusive deep inelastic scattering of polarised positrons from
polarised targets of H, D, and $^3$He.  These data give information on
the spin structure of the nucleon.  This paper describes the forward
angle spectrometer built for this purpose.  The spectrometer includes
numerous tracking chambers (micro-strip gas chambers, drift and
proportional chambers) in front of and behind a 1.3~T.m magnetic
field, as well as an extensive set of detectors for particle
identification (a lead-glass calorimeter, a pre-shower detector, a
transition radiation detector, and a threshold \v{C}erenkov detector).
Two of the main features of the spectrometer are its good acceptance
and identification of both positrons and hadrons, in particular
pions.  These characteristics, together with the purity of the
targets, are allowing HERMES to make unique contributions to the
understanding of how the spins of the quarks contribute to the spin of
the nucleon.
\end{abstract}

\ \

\normalsize
{\em (Submitted to Nuclear Instruments and Methods)}

\end{center}
\twocolumn

\section{\bf Introduction.}

The HERMES experiment (\underline{HER}A \underline{ME}asurement of
\underline{S}pin) is a second generation polarised deep inelastic
scattering (DIS) experiment to study the spin structure of the
nucleon. It is being run at the HERA storage ring at
DESY\footnote{HERA is an e-p collider but only the e beam is used by
HERMES}.

Several experiments over the last decade have provided accurate data
on the polarisation asymmetry of the cross-section for {\em inclusive}
scattering where only the scattered lepton is detected.  These
experiments have been interpreted as showing that at most 30\% of the
nucleon spin comes from the spins of the quarks.  Further knowledge of
the origin of the nucleon's spin can be gained by studying {\it
semi-inclusive} processes involving the detection of hadrons in
coincidence with the scattered lepton.  These data offer a means of
`flavour-tagging' the struck quark to help isolate the contributions
to the nucleon spin of the individual quark flavours, including the
sea quarks.  The interpretation of semi-inclusive data is made clearer
if the type of hadron is identified. This is a central theme of the
HERMES experiment, which can identify pions and will add kaon
identification for 1998.

The physics program for HERMES is very broad. The experiment
contributes inclusive data with qualitatively different systematic
uncertainties to improve the world data set for the {\em x} dependence
and the integral of the spin structure function g$_1$($x$) for both
the proton and the neutron (see below for the definition of $x$).
Most importantly, HERMES is providing new precise data on
semi-inclusive processes by virtue of the good acceptance of the
spectrometer combined with hadron identification and the purity of the
targets.

The HERA storage ring can be filled with either electrons or positrons,
which are accelerated to 27.5~GeV.  Positrons have been used since
1995 because longer beam lifetimes are then possible.  Since with few
exceptions such as the luminosity measurement, the physics processes
are the same for positrons and electrons, the term ``positron'' will
be used for both in this paper.

In addition to studying polarised DIS using $\rm \vec{H}, \vec{D}, and
\, \vec{^3He}$ targets, data are collected with unpolarised gases
($\rm H_2, D_2, ^3He, N_2$).  This provides in a relatively short
period of time (2-3 weeks) high statistics data sets that are used to study
important properties of the nucleon not related to spin, such as the
flavour asymmetry of the sea, as well as hadronisation in nuclei.

\section{\bf General Description.}

\begin{figure*}
\hskip 0.01cm \psfig{figure=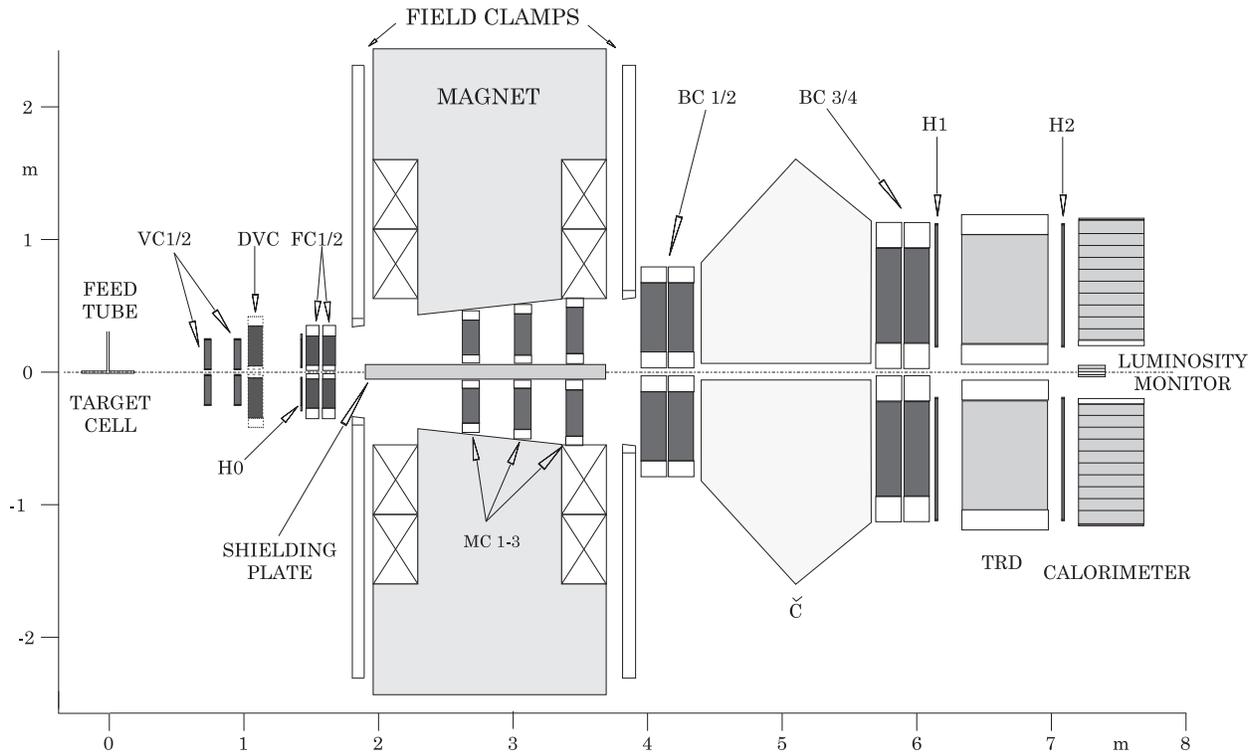,width=16.5cm}
\caption{Schematic side view of the HERMES spectrometer. See the text
for the meaning of the labels.
\label{fig:spect}}
\end{figure*}

The HERMES experiment is located in the East hall of the HERA storage
ring complex at DESY.  The spectrometer is a forward angle instrument
of conventional design. It is symmetric about a central, horizontal
shielding plate in the magnet.  Due to this symmetry, the description
of the detectors contained in this paper will apply typically to only
one half of the spectrometer, in particular with respect to the number
of detectors quoted and their dimensions.  A diagram of the
spectrometer is shown in Fig.~\ref{fig:spect}.  The coordinate system
used by HERMES has the z axis along the beam momentum, the y axis
vertical upwards, and the x axis horizontal, pointing towards the
outside of the ring.  The polar ($\theta$) and azimuthal ($\phi$)
scattering angles as well as the initial trajectory for the
determination of the particle's momentum are measured by the front
tracking system, which consists of microstrip gas chambers (MSGC,
referred to as the vertex chambers (VC)) and drift chambers (DVC,
FC1/2).  The momentum measurement is completed by two sets of drift
chambers behind the magnet (BC1/2 and BC3/4).  In addition, there are
three proportional chambers in the magnet (MC1/3) to help match front
and back tracks as well as to track low momentum particles that do not
reach the rear section of the spectrometer.

Particle identification (PID) is provided by a lead-glass calorimeter,
a pre-shower detector (H2) consisting of two radiation lengths of lead
followed by a plastic scintillator hodoscope, a transition radiation
detector (TRD) consisting of six identical modules, and a threshold
\v{C}erenkov detector (\v{C}).  The particle identification system was
designed to provide a hadron rejection factor of 10$^4$ to yield a
very clean DIS positron sample, and exceeds this in practice.  Pion
identification is provided by a threshold \v{C}erenkov detector.  An
upgrade to a ring imaging \v{C}erenkov detector that will identify
both pions and kaons is being prepared for operation in 1998.  The
calorimeter and pre-shower detector are included in the trigger along
with a second hodoscope (H1) placed in front of the TRD.  An
additional trigger hodoscope (H0) was included in 1996 in front of
FC1 to reduce the trigger rate from background caused by the proton
beam.

The acceptance is limited at small angles by an iron plate in the beam
plane, which shields the positron and proton beams from the magnetic
field of the spectrometer magnet.  Both beams go through the
spectrometer, separated by 72~cm.  Particles with scattering angles
within $\pm$170~mrad in the horizontal direction and between
+(--)40~mrad and +(--)140~mrad in the vertical direction are accepted.
Therefore, the range of scattering angles is 40~mrad to 220~mrad.
Spin structure functions depend on $x$ ($x$=~Q$^2$/2M$\nu$ where Q is
the four-momentum transfer in the DIS reaction, $\nu$ is the energy
transfer, and M is the mass of the nucleon). The variable $x$ can be
interpreted as the fraction of the nucleon's momentum carried by the
struck quark. The $x$ range covered by the HERMES experiment is
0.02-1.0, although there is little count rate for x$\geq$0.8 when W
$>$ 2 GeV (W is the photon-nucleon invariant mass).

The experiment is mounted on a large platform that can move on rails
together with an attached trailer (ET) containing the electronics and
the gas systems.  A fixed but removable concrete wall between the
platform and the ET shields the main part of the hall from radiation
and hence allows access to the electronics while the accelerators are
running.  Cables and gas pipes between the detectors and the ET are
routed beneath the platform to a large cable tray passing under the
shielding wall.  The experiment was assembled in the East hall outside
the shielding during the 1994 HERA run.  After the shielding wall had
been dismantled, the platform/trailer was moved into place in the
ring.  The shielding wall was re-erected between the platform and the
electronics trailer in such a way that the experiment could be moved
far enough within the shielding to allow the HERA maintenance tram
access to the tunnel on 24 hours notice.  

\section{The Target Region.}

The HERMES target consists of gas from a polarised source fed into a
storage cell internal to the HERA positron ring.  In 1995, an
optically pumped polarised $\rm \vec{^3He}$ cell was used to supply
gas to the storage cell \cite{TAR-3he}, while an atomic beam source
(ABS) was used in 1996-97 to produce a $\rm \vec{H}$ target
\cite{TAR-ABS}.  The ABS will be modified for operation with $\rm
\vec{D}$ in 1998-99.  It is also possible to inject unpolarised gases
into the storage cell.  The following were used in the first three
years of operation: $\rm H_2, D_2, ^3He, N_2$.

\begin{figure}
\hskip 0.01cm \psfig{figure=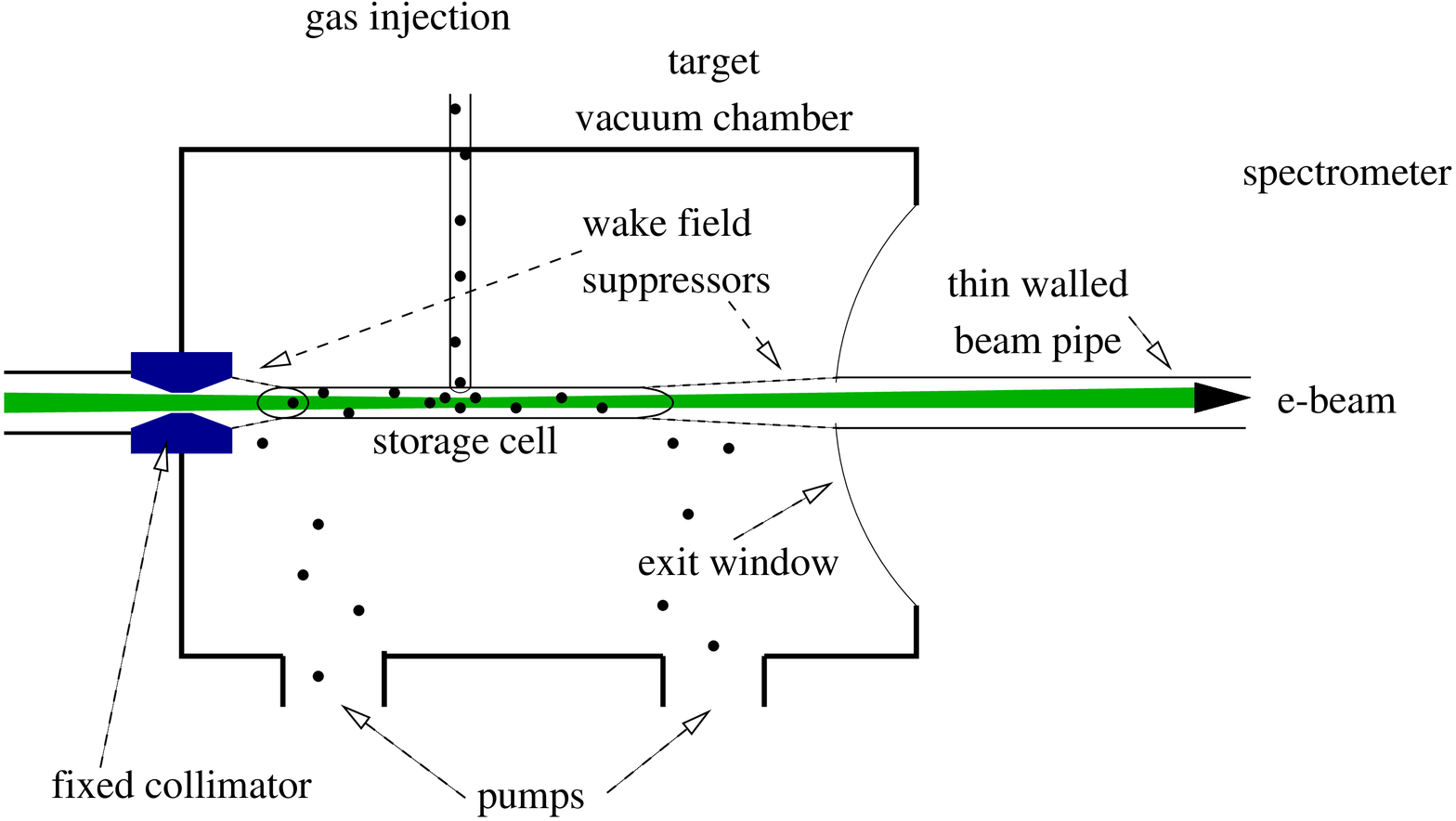,width=3.10in}
\caption{Schematic of the target region.  See the text for more
explanation. 
\label{fig:target}}
\end{figure}

The target region is shown schematically in Fig.~\ref{fig:target}.
The gas enters an open-ended (ie. windowless) T-shaped tube that
confines the gas atoms in a region around the positron beam.  The
storage cell increases the areal target density by about two orders of
magnitude compared to a free atomic beam.  The gas atoms leak out the
open ends of the target cell and are pumped away by a high speed
differential pumping system.  In this way, the number of atoms seen at
the target location by the beam is maximized while minimizing the
effect on the stored beam.  The storage cell is an elliptical tube,
9.8~mm high by 29.0~mm wide and 400~mm in length.  It was made of
125~$\mu$m thick ultra-pure aluminium in 1995 and 75~$\mu$m aluminium
in 1996-97.  The transition from the cell to the beam pipe was made
smooth to avoid the generation of wake fields that could cause heating
and increase the emittance of the beam.  This was accomplished using
thin perforated tubes called ``wake field suppressors''.

The amount of synchrotron radiation created upstream of the HERMES
area is considerable, even though it is reduced by the addition of two
weak dipoles downstream of the last bending magnet in the arc of the
accelerator. If it were allowed to hit the target cell, it would
damage the cell coating, and on the order of 10$^{13}$~-~10$^{14}$
soft synchrotron photons per second would Compton scatter into the
spectrometer acceptance. This is prevented by a system of two
collimators installed near the target.  C1 is placed 2~m upstream of
the target and protects the next collimator from direct radiation.  It
actually consists of two moveable collimators, C1H in the horizontal
direction and C1V in the vertical, separated by 0.5~m.  These
collimators are opened during beam injection and operated at a narrow
but safe setting (15$\sigma$ + 0.5~mm) during HERMES operation.
During the first three years of operation, typical beam parameters
were $\sigma_x$=~0.31~mm and $\sigma_y$=~0.07~mm.  C2 has a fixed
aperture that is slightly larger than C1, and is located right in
front of the storage cell to protect it from photons scattered in C1.
It also protects the cell from direct radiation during beam injection.
Note that only C2 is shown in Fig.~\ref{fig:target}.

Particles scattered into the spectrometer acceptance exit the target
chamber through a thin (0.3~mm) stainless steel foil.  Background from
showers (initiated mostly by $\gamma$'s from $\pi^0$ decay) is reduced
by also making the beam pipe just downstream of the target chamber of
thin stainless steel.

\section{The Magnet.}

The HERMES spectrometer magnet is of the H-type with field clamps in
front as well as behind in order to reduce the fringe fields at the
position of the drift chambers FC2 and BC1. A massive iron plate in
the symmetry plane shields the positron and proton beams as they
pass through the magnet.

\noindent
The most important features of the magnet are:
\begin{itemize}
\item{It is capable of providing a deflecting power of
      $\int Bdl$ = 1.5~T$\cdot$m, although it is operated
      at 1.3~T$\cdot$m to reduce power consumption.  The
      smaller field integral does not significantly impair the
      performance of the spectrometer.
      The variation of the deflecting power within the acceptance is
      less than 10 \%.}
\item{The gap between the pole faces encloses the geometrical
      acceptance of $\pm$(40..140)~mrad in the vertical direction.
      In the horizontal direction $\pm$170 mrad plus another $\pm$100~mrad
      starting halfway through the magnet is provided.  The pole faces
      are tilted parallel to the limits of the vertical angular acceptance.
      Due to the limited space of 8.5~m for the HERMES spectrometer between
      the center of the target and the first quadrupole magnet of the
      electron machine, only 2.2~m between the drift chambers FC2 on the
      front and BC1 on the rear side are utilized for the magnet.}
\item{The fringe field at the positions of the adjacent drift chambers
      does not exceed 0.1~T.}
\item{An effective magnetic shielding substantially reduces the effect 
      of the magnet on the proton and positron beams.  In particular, 
      the sextupole moment of the magnetic field 
      in the beam tubes is minimized.} 
\item{A correction coil with a deflecting power of 0.08~T$\cdot$m is
      accommodated inside the shielding of the positron beam pipe. 
      This coil is used to correct for fringe fields and the 
      imperfections of the magnetic shielding in this section of the iron 
      plate.  It is also intended to compensate the transverse holding 
      field of the target when operating with transverse polarisation.  
      It serves as an element of a closed orbit bump with no net global 
      effect on the positron beam.}
\end{itemize}

\noindent
Magnetic model calculations were done with the 3D-codes MAFIA
\cite{MAG-mafia} and TOSCA \cite{MAG-tosca}.  The calculations agree
with subsequent magnetic measurements within a few percent. 
The field measurements were done with a 3D-Hall probe on an automated
3D-mapping machine. The results exhibit an overall reproducibility of about
10$^{-3}$, and a field pattern as expected.  No homogeneous regions are
found due to the high gap to length ratio of the magnet,
and a pronounced step in the field is observed
that reflects the structure of the iron shielding
plate.  The detailed field map was integrated into the track
reconstruction algorithm, as described later.

\section{The Tracking System}

The tracking system serves several functions, which 
vary somewhat in their performance requirements: 

\begin{itemize}

\item Determine the event vertex in the target cell to ensure
that reconstructed events come only from the sub-mm beam envelope in the
target gas, with the expected z-distribution. Also, the analysis of
events with multiple tracks requires either a common vertex, or
possibly an additional displaced vertex for a reconstructed particle
decay. 

\item Measure the scattering angles for kinematic reconstruction.

\item Measure the particle momentum from the track deflection in the
spectrometer dipole magnet.

\item Identify the hits in the PID detectors associated with each 
track.

\end{itemize}

When all tracking detectors are fully operational, the momentum
resolution for positrons is limited by Bremsstrahlung in the material
of the target cell walls, the 0.3~mm thick stainless steel vacuum
window, as well as the first tracking detectors. Even the resolution
in scattering angle is normally limited by multiple scattering in this
material. An exception was 1995, the first year of operation, when the
low efficiency of most planes of the Vertex Chambers (improved
for 1996 for the top detector) made it more attractive to reconstruct
the tracks using only the drift chambers (DC), as will be explained
later in more detail. This mode of operation resulted in the
resolution of reconstructed kinematic quantities being much more
sensitive to the DC spatial resolution.

The locations of the tracking detectors are shown in Fig.~\ref{fig:spect} 
and listed in Table \ref{trtab}, which also summarizes their properties.

The microstrip gas counters VC1/2, the drift vertex chambers DVC's,
and the front drift chambers FC1/2 provide both vertex reconstruction
to the target, and definition of the scattering angle. In conjunction
with the front tracking, the back drift chambers BC1/2 and BC3/4
measure the magnetic deflection and hence the momentum. The BC's also
identify the cells in the PID detectors associated with each
track. The proportional chambers inside the magnet (MC1/3) were
originally intended to ensure that multi-track ambiguities could be
resolved. As it happens, chamber occupancies are low enough that this
can be accomplished using the drift chambers alone. However, the MC's
are found to be very useful for momentum analysis of low energy decay
products that are deflected too much to reach the downstream tracking
detectors.

\begin{table*}
\caption{Tracking Chamber properties.}
\label{trtab}
\begin{center}
\begin{sideways}
\begin{tabular}{|l|cc|c|cc|ccc|cc|}     \hline
 CHAMBER  &\multicolumn{2}{c|}{Vertex}  & Drift Vertex &
\multicolumn{2}{c|}{Front} & \multicolumn{3}{c|}{Magnet} &
\multicolumn{2}{c|}{Back}   \\ \hline
Detector name & VC1 & VC2 & DVC & FC1 & FC2 & MC1 & MC2 & MC3 & BC1/2 & BC3/4 \\ 
mm from target & 731 & 965 & 1100 & 1530 & 1650 & 2725 & 3047 & 3369 &
 4055 & 5800 \\
Active Area: & & & & & & & & & & \\
~~Horizontal (mm) & 323 & 393 & 474 & 660 & 660 & 996 & 1210 & 1424 & 1880 & 2890 \\ 
~~Vertical (mm) & 137 & 137 & 290 & 180 & 180 & 263 & 306 & 347  & 520  & 710 \\ 
Cell design & \multicolumn{2}{c|}{microstrip gas} &
horizontal-drift & \multicolumn{2}{c|}{horizontal-drift} &
\multicolumn{3}{c|}{MWPC} & \multicolumn{2}{c|}{horizontal-drift} \\ 
Cell width (mm) & \multicolumn{2}{c|}{0.193}  & 6 & \multicolumn{2}{c|}{7} &
\multicolumn{3}{c|}{2} & \multicolumn{2}{c|}{15} \\ 
A-C plane gap (mm) & \multicolumn{2}{c|}{3}  & 3 & \multicolumn{2}{c|}{4} &
\multicolumn{3}{c|}{4} & \multicolumn{2}{c|}{8} \\ 
Anode (A) material& \multicolumn{2}{c|}{200\,$\mu$m glass(Al)} &
W(Au) & \multicolumn{2}{c|}{W(Au)} &
\multicolumn{3}{c|}{W(Au)} & \multicolumn{2}{c|}{W(Au)} \\ 
Anode wire diameter & \multicolumn{2}{c|}{7\,$\mu$m} &
30~$\mu$m & \multicolumn{2}{c|}{20\,$\mu$m} &
\multicolumn{3}{c|}{25\,$\mu$m} & \multicolumn{2}{c|}{25\,$\mu$m} \\ 
Potential wire mat$'$l & \multicolumn{2}{c|}{Al strip} &
Be-Cu(Au) & \multicolumn{2}{c|}{Al(Au)} &
\multicolumn{3}{c|}{} & \multicolumn{2}{c|}{Be-Cu(Au)} \\ 
Potential wire dia.  & \multicolumn{2}{c|}{85\,$\mu$m} &
50\,$\mu$m & \multicolumn{2}{c|}{50\,$\mu$m} &
\multicolumn{3}{c|}{} & \multicolumn{2}{c|}{127\,$\mu$m} \\
Cathode (C) material& \multicolumn{2}{c|}{Al on glass} &
Al on Mylar & \multicolumn{2}{c|}{Al on Mylar} &
\multicolumn{3}{c|}{Be-Cu wires} &
\multicolumn{2}{c|}{C on Kapton} \\ 
Cathode thickness & \multicolumn{2}{c|}{200\,$\mu$m} &
34\,$\mu$m &  \multicolumn{2}{c|}{6.4\,$\mu$m} &
\multicolumn{3}{c|}{90\,$\mu$m @ 0.5~mm pitch} &
\multicolumn{2}{c|}{25.4\,$\mu$m} \\ 
Gas composition: & \multicolumn{2}{c|}{DME/Ne} &
Ar/CO$_{2}$/CF$_{4}$ & \multicolumn{2}{c|}{Ar/CO$_{2}$/CF$_{4}$} &
\multicolumn{3}{c|}{Ar/CO$_{2}$/CF$_{4}$} &
\multicolumn{2}{c|}{Ar/CO$_{2}$/CF$_{4}$} \\ 
(\%) & \multicolumn{2}{c|}{50/50} &
90/5/5 & \multicolumn{2}{c|}{90/5/5} &
\multicolumn{3}{c|}{65/30/5} &
\multicolumn{2}{c|}{90/5/5} \\ 
U,V stereo angle & \multicolumn{2}{c|}{$+5^{\circ}$,~$-90^{\circ}$}  &
$\pm30^{\circ}$ & \multicolumn{2}{c|}{$\pm30^{\circ}$} &
\multicolumn{3}{c|}{$\pm30^{\circ}$} & \multicolumn{2}{c|}{$\pm30^{\circ}$} \\
Resolution/plane ($\sigma$)& \multicolumn{2}{c|}{65\,$\mu$m} &
220\,$\mu$m & \multicolumn{2}{c|}{225\,$\mu$m} &
\multicolumn{3}{c|}{700\,$\mu$m} & 275\,$\mu$m & 300\,$\mu$m \\
Wires in X plane & 1674 & 2046 & 80 & 96 & 96 & 496 & 608 & 720 & 128 & 192 \\
Wires in U,V plane & 2170 & 2170 &
96 & 96 & 96 & 512 & 608 & 720 & 128 & 192 \\
Module config. & VUX & XVU &
XX$'$UU$'$VV$'$ & \multicolumn{2}{c|}{UU$'$XX$'$VV$'$} &
\multicolumn{3}{c|}{UXV} & \multicolumn{2}{c|}{UU$'$XX$'$VV$'$} \\
Rad. length/module & \multicolumn{2}{c|}{0.8\%} &
0.25\% & \multicolumn{2}{c|}{0.075\%} &
\multicolumn{3}{c|}{0.29\%} & \multicolumn{2}{c|}{0.26\%} \\
Number of modules & 1 & 1 & 1 & 1 & 1 & 1 & 1 & 1 & 2 & 2 \\ 
(upper or lower) & & &~~~~~~~~~ & & & & & ~~~~~~~~~ & & \\
Channels/module & 6014 & 6386 & 544 & 576 & 576 & 1520 & 1824 & 2160 &
768 & 1152 \\ \hline
Total channels & \multicolumn{2}{c|}{24800} & 1088 & \multicolumn{2}{c|}{2304} &
\multicolumn{3}{c|}{11008} & \multicolumn{2}{c|}{7680} \\ \hline
\end{tabular}
\end{sideways}
\end{center}
\end{table*}

The horizontal length of the BC's precluded the use of long horizontal
wires for y measurements. Hence all planes are one of three types with
wires oriented either in the vertical ( for X measurement), or tilted
30$^\circ$ right or left (U and V planes). In order to allow the use of
fast reconstruction algorithms, all tracking detectors are restricted
to this geometry. Like the rest of the spectrometer, the tracking
system is symmetric about the beam plane. 

\subsection{Vertex Chambers}

The purpose of the vertex chambers is to provide high-precision
measurements of the scattering angle and the vertex position over the
full acceptance of the experiment, in the presence of a significant
background flux. The proximity of the VC to the target implies that at
the minimum scattering angle, the tracks from the downstream end of
the target cell pass the VC active area 20~mm from the beam axis, or
only 5~mm from the beam tube. In combination with the need to minimize
high-Z materials near the beam tube that could convert the numerous
photons from $\pi^0$ decay into high multiplicity showers, this
geometry presented severe problems in mechanical design that would
have been very difficult to solve in the context of wire chamber
frames. All these requirements are met through the use of a relatively
new technology, a microstrip gas chamber (MSGC). In this technique
\cite{VC-msgc}, metal strips are etched on a thin, semi-rigid
substrate of high resistivity, allowing for higher drift fields and
much smaller cell pitch than in a conventional drift chamber. Each of
the upper and lower VC assemblies contains six MSGC planes, grouped
into two ``modules" (VUX and XUV for VC1 and VC2 respectively), which
are sufficiently separated to provide a precise determination of the
front track by the VC alone. The total radiation length of one module
is 0.8\% X$_0$ and the thickness of the target chamber vacuum window
is 1.7\% X$_0$, so that for momenta below about 7 GeV the angular
accuracy becomes limited by multiple scattering. An inherent cost
associated with the MSGC choice is the large channel count -- 24,800
channels in total. This requires a highly integrated readout system,
including analog electronic signal storage on the planes and
multiplexed digitization at the detector.

\begin{figure}
\hskip 1.0cm \psfig{figure=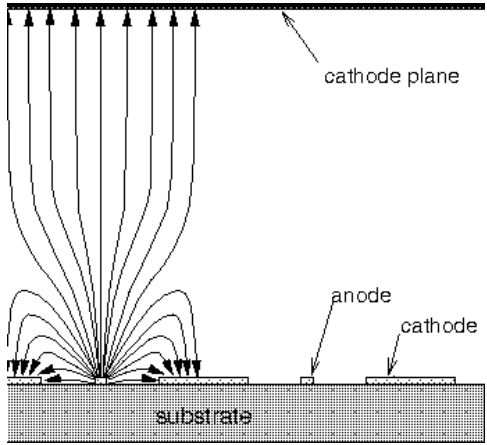,width=2.5in}
\caption{Schematic view of a drift cell in a MicroStrip Gas Counter,
showing the field lines. This
figure has been taken from ref.\cite{VC-vdMarel}. 
\label{fig:msgcell}}
\end{figure}

The basic features of the MSGC drift cell are shown in 
Fig.~\ref{fig:msgcell}.  Ionization electrons created by a traversing
particle in a 3 mm thick gas volume move in a drift field of about 5
kV/cm directly away from a planar cathode and towards a 200 $\mu$m
thick glass substrate that carries a pattern of fine etched aluminum
strips. The field is produced by a 1800~V potential on the cathode
planes.  A voltage difference of 580~V is applied between the 7 $\mu$m
wide anode strips and the 85~$\mu$m wide cathode strips forming the
pattern on the substrates.  As a result, gas multiplication occurs in
the vicinity of the anode, a process comparable to that in a wire
chamber.  High gas amplification at 580V is obtained by using a
DiMethyl-Ether(DME)/Ne mixture (50/50\%).  The gas gain is set at a
level such that the signals are about three times that of silicon
detectors of the same material thickness, leading to a better signal
to noise ratio. Even with only single bit per cell digital read-out, a
spatial resolution much better than the cell pitch is achieved, due to
charge sharing between cells.

The metallic strip pattern that creates the electrical field
configuration is carried on substrates, the material of which is a
critical choice. Low Z is preferred to minimize the effects of
multiple scattering and Bremsstrahlung. Surface properties are
important as electrons, ions, and gas pollutants may stick and cause
field modifications, discharges, ageing, and radiation damage. Since
part of the avalanche falls on the substrate instead of on the
conductive strip, the substrate should have a certain degree of
conductivity in order to remove the collected surface charge.
D-263 glass was chosen as the  substrate material~\cite{VC-D263}.
It consists of SiO$_x$ (64.1~\%), Al$_2$O$_3$ (4.5~\%), K$_2$O (6.6~\%), 
TiO (4.2~\%), ZnO (6.1~\%), Na (6.1~\%), with the remainder (8.4~\%) 
unspecified. It has a bulk resistivity of 10$^{15}$~$\Omega$cm,
equivalent to a surface resistivity of 10$^{17}$~$\Omega/\Box$,
caused by the migration of electrons and Na$^+$ ions. The thickness of
the substrates is 0.2 mm, the minimum required for safe handling and
to have sufficient rigidity to resist electrostatic distortions. 

The active area of the U or V planes is as large as
484$\times$137~mm$^2$. The largest substrate area that could be
processed is 150 $\times$ 200 mm$^2$. Therefore, 5 substrates are combined
to form one 30$^\circ$ plane. The substrates are cut along the cathode
strip, and glued on a carrier frame, leading to a gap of on average
461~$\mu$m between the anode wires of consecutive substrates.  No loss
in efficiency at these boundaries has been observed.

The PC carrier frame to which the substrates are glued has sufficient
mechanical stiffness to be self-supporting. The carrier frames
are mounted in a fixture attached to the inside of the lid of the box
that acts as the gas enclosure. The fixture provides a means of
precisely aligning the planes with respect to fiducials on the outside
of the lid. The wall of the box, which is shaped to curve around the
beam tube, is only 0.5\,mm thick and made of low-Z material. The thin
windows for the detected particles have two layers enclosing
non-flammable flush gas to ensure no leaks of flammable gas. 

In order to minimize external data flow, digitization of the signals
is done in an electronic assembly mounted on the detector. The high
degree of digitizer multiplexing required for economy is made possible
by continuously storing the charge buckets collected by each anode
strip in a switched-capacitor analog pipeline. There are 64
preamplifiers and pipelines, each 32 steps deep, integrated on each
Analog Pipeline Chip (APC). The pipelines are clocked with
the HERA bunch frequency (10.4~MHz). If an event trigger occurs any
time within 3.1~$\mu$s (32 $\times$ 96~ns) after an event, the
pipelines are stalled before the data are shifted out and lost. The
pipelines are advanced until the buckets of interest in the vicinity
of the bunch causing the trigger are about to appear at the pipeline
output stage. Then the pipelines are shifted much more slowly as the
buckets of interest are presented to the ``tracking discriminators",
which were designed to compensate the rising baseline of the APC
chip. A sample and hold circuit is used to keep the threshold at a
fixed value above the baseline offset, given by the readout value of
the previous channel. If a hit is detected, the baseline restore
operation is suspended, until the following channel no longer exceeds
the discriminator level. For that reason, the system is limited to
handle a maximum cluster size of up to four neighboring anode
strips. The first two channels of each chip are not connected to an
anode strip, and serve as a reference for the baseline
restoration. The data from each discriminator (62 channels) are stored
in its shift register. As soon as all digital data have been stored,
the inputs of the APC chips are enabled and a new trigger can be
accepted. The dead time during readout of the APC chips and digitization
into the shift register is 56~$\mu$s. Simultaneously, serial readout
of the shift registers is started. Addresses of signals above
threshold are loaded into memory stacks. If during this operation a
new trigger arrives, the new event waits in the APC pipeline until the
data of the previous event are transferred into the memory
stacks. Subsequently, the memory stacks are read out serially via a
data driver into a Struck ECL300 FASTBUS module.

The efficiency of the vertex detector is measured in two ways. In the
internal tracking approach, tracks are identified using VC data alone,
requiring hits in five of six planes. The track is reconstructed from
four planes, and a hit is accepted in the fifth plane if it is within
200 $\mu$m of the track. Only tracks are considered that point to the
target and are reconstructable by the rest of the tracking system. The
efficiency of the fifth plane is then defined as the fraction of
events with a hit close to the track. In the external tracking
approach, the efficiency is determined with tracks reconstructed using
the drift chamber system, extrapolated to the vertex chamber. In both
cases, efficiencies in the range 60-90\% were found for the 1995 data,
where a prototype version of the APC chip was used that exhibited a
strongly non-linear baseline. This caused high noise. Due to the
fragility of some of the anode resistors that were etched on the
substrates, the high voltage could not be raised sufficiently to get a
high enough signal-to-noise ratio. For 1996 running, the substrates
and APC chips of the upper chamber were replaced, resulting in much
improved efficiencies of around 95\%.  The lower half was similarly
upgraded for 1997 running.  The resolution has been determined by
comparing the distance between the track, formed with 5 planes, and
the hit in the sixth plane, and also by comparing the horizontal angle
of the track calculated from the X planes with that from the U and V
planes. Both methods yield a resolution of about 65~$\mu$m ($\sigma$)
per plane.

\begin{figure}
\hskip 0.01cm \psfig{figure=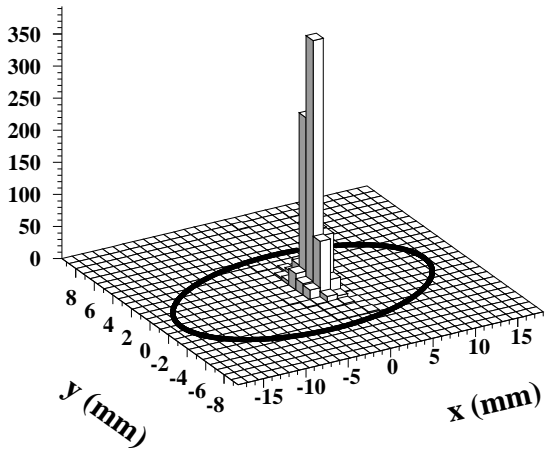,width=8.0cm}
\caption{Vertex reconstruction in the transverse coordinates using the VC 
alone.  This plot was made using two-track events only to get a well
defined vertex.  The outline of the target cell is shown as the ellipse.
\label{fig:vc-vertex}}
\end{figure}

The vertex reconstruction capability of the VC is illustrated in
Fig.~\ref{fig:vc-vertex}, which is a histogram of the transverse
coordinates of the point of closest approach of two tracks in an
event.  A well defined beam envelope is seen.  Furthermore, the figure
shows that the vertices are well separated from the target cell wall
represented by the ellipse.

Over three years of operation, there has been no observed degradation
of performance that would suggest ageing effects.

\subsection{The Drift Chambers}

The drift chambers DVC, FC1/2, BC1/2 and BC3/4 are of the
conventional horizontal-drift type. Each layer of drift cells consists
of a plane of alternating anode and cathode wires between a pair of
cathode foils. The cathode wires and foils are at negative high
voltage with the anode sense wires at ground potential. The chambers
are assembled as modules consisting of six such drift cell layers in
three coordinate doublets ($UU^\prime$, $XX^\prime$ and $VV^\prime$).
The wires are vertical for the $X$ planes and at an angle of
$\pm$30$^\circ$ to the vertical for the $U$ and $V$ planes. The
$X^\prime$, $U^\prime$ and $V^\prime$ planes are staggered with
respect to their partners by half the cell size in order to help
resolve left-right ambiguities. Each sense plane consists of a
fiberglass-epoxy frame laminated to a printed circuit board (PCB)
that has solder pads for the wires. Except in the case of the DVC's,
the PCB's carry the traces leading past O-ring gas seals to
external connectors.  The opposite ends of the cathode wires connect
to internal distribution buses for the negative high voltage.  Each
module is enclosed between metallized Mylar gas windows mounted on
metal frames. Identical modules are situated above and below the beam
pipe.

The frames of all the drift chambers except the DVC's have sufficient
length to allow both ends of all wires in the $U$ and $V$ planes to
terminate on the long edges of the frames. Hence all $U$ and $V$ wires
and their PCB traces have the same length and similar capacitance. The
onboard electronics are mounted along only the one long edge opposite
to the beam pipes. This arrangement does not sacrifice mechanical
stiffness of the frame. Since the inside aperture of each frame is
only as large as the active area in common with all the frames, each
stressed frame member is as short as possible. A shallow relief is
milled in the inactive end regions of the $U$ and $V$ frames to allow
the wires to stand off from the frame material. Proximity to the beam
of the active areas of all chambers required external 
notches to be machined in the frames to accommodate the beam pipes. 
As these notches are short, they do not seriously compromise the 
mechanical stiffness of the frames.

The choice of gas mixture for the drift chambers was governed by the
serious inconvenience of controlling the hazards of a flammable gas in
a tunnel environment. As well, there is an advantage in limiting the
drift cell occupation time and hence the number of extraneous hits
that must be accommodated in track finding. The mixture
Ar(90\%)/CO$_2(5\%)$/CF$_4(5\%)$ is both fast and non-flammable
\cite{DC-gas}. Its drift velocity has been measured as a function of
field - eg. $>$ 7~cm/$\mu$sec at E=800~V/cm \cite{DC-drift}. FC2 and
BC1 operate in a dipole fringe field that reaches 0.1~T.  Mixtures
containing CF$_4$ have been shown to result in compromised resolution
due to electron attachment \cite{DC-Biagi}, and large Lorentz drift
angles in magnetic fields.  However, they have the advantages of short
occupation time and long chamber lifetime~\cite{DC-life}.

The gas pressure and flow through all drift chambers is regulated by a
system of valves and mass flow controllers, all under the supervision
of an industrial programmed controller. A large fraction of the flow
(80-90\%) is recirculated through purifiers that remove oxygen and
water vapour, but not nitrogen. In 1995, nitrogen from gas leaks in
the drift chambers and from diffusion in the window foils accumulated
in the loop to an equilibrium concentration of 1.1\%. This influenced
the drift velocity in the gas mixture and resulted in a gain reduction
as confirmed by independent studies \cite{DC-schmidt}. Work on the
chamber gas leaks as well as decreasing the re-cycled fraction from
90\% in 1995 to 80\% in 1996 (20\% exhaust), reduced the nitrogen
contamination to a stable level of 0.4\% with negligible influence on
both drift velocity and gain.

The DC readout system consists of Amplifier/Shaper/Discriminator (ASD)
cards mounted onboard the drift chambers, driving ECL signals on 30~m
long flat cables to LeCroy 1877 Multihit FastBus TDC's in the external
electronics trailer. The design goal (especially in the case of the FC's
and DVC's) was chamber operation at low gas gain ($\sim 10^4$) to maximize
chamber lifetime and reduce particle flux dependence of the gas gain
due to space charge effects. To combine this goal with good spatial
resolution requires operation at low threshold -- well below $10^5$
electrons from the wire. This in turn requires both low loise
performance of the amplifiers as well as great care in the
radio-frequency design of the electronics enclosure on the chambers,
to maintain electronic stability. All ground elements including the
ASD cards themselves should be considered as radio-frequency
resonators, and the lowest frequency of any resonant mode should be
kept well above the pass-band of the amplifiers. For example, this
dictates that the ground plane of each 16-channel ASD card
continuously contact the electronics enclosure along both sides of the
card via special high-conductivity card guides. 

During the first year of operation in 1995, all drift chambers
performed satisfactorily and alone provided the tracking data that were
used in the physics analysis.  As discussed later, performance was
improved in 1996-97 by the addition of the DVC's.

\subsubsection{The Front Drift Chambers.}

The front drift chambers FC1 and FC2 provide good spatial resolution
immediately in front of the spectrometer magnet. Each chamber above or
below the beam consists of one module of six sense planes; thus the
total front drift chamber package presents twelve sense planes to a
charged particle.  Design parameters are listed in Table \ref{trtab}.

A drift cell size of only 7 mm (3.5 mm maximum drift length) was
chosen to provide fine spatial granularity to maximize tolerance to both
correlated and uncorrelated background. It emerged that uncorrelated
background is mostly from collimator showers, at the average level of 
only 0.1 tracks per event. For a minor fraction of the events, the FC
occupation from correlated tracks can be substantial, but is always
acceptable. 

The ASD design is similar to that of the LeCroy 2735DC card, except
with a low-noise front end and pulse shaping added.  With the input
transistor (MMBR941LT1) used in common emitter mode with feedback,
channel to channel gain variation is about $\pm10\%$.  In order to
optimize pulse pair resolution and hence rate capability, pulse
shaping between stages is also added.  A long tail due to positive
ions from the avalanche drifting to the cathodes is cancelled by
shaping with two stages of pole-zero-cancelled differentiation.  To
facilitate testing of the chamber system, a linear monitor point is
provided for each channel.

The threshold can be varied from approximately 0.07 to 1.25$\,\mu$A
at the input. At the lowest value, it will trigger efficiently on an
impulse charge of approximately 12,000 electrons, resulting
in 8$\,$ns wide pulses at both the linear monitor and ECL logic
outputs. An impulse charge of 1600 electrons produces a linear pulse of
about the same magnitude as the RMS noise. In the 1995 running period,
the FC threshold was dictated by ambient noise radiated by the target
r.f. system. In 1996, the target r.f. noise was reduced
and the electronics
enclosure was improved, allowing a threshold of 140$\,$nA from
the wire, corresponding to a charge impulse of about 20,000 electrons.
The power consumption per board is 0.3$\,$A at +5$\,$V and 0.55$\,$A
at $-5.2\,$V, or 270$\,$mW per channel. Hence moderate forced air
cooling is provided.

\begin{figure} 
\hskip 1.0cm \psfig{figure=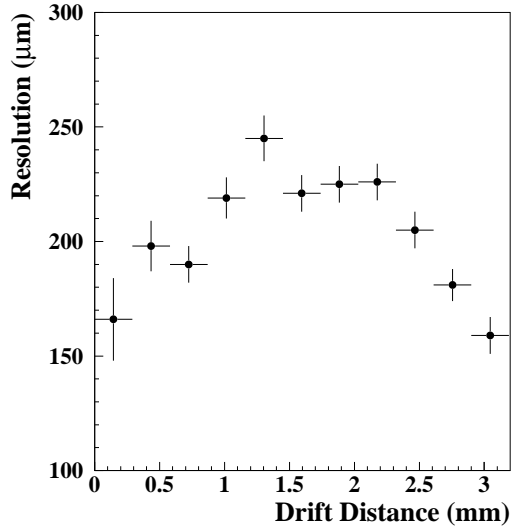,height=3.0in}
\caption{FC resolution as function of drift distance.
\label{fig:fc-res}}
\end{figure}

\begin{figure} 
\hskip 1.0cm \psfig{figure=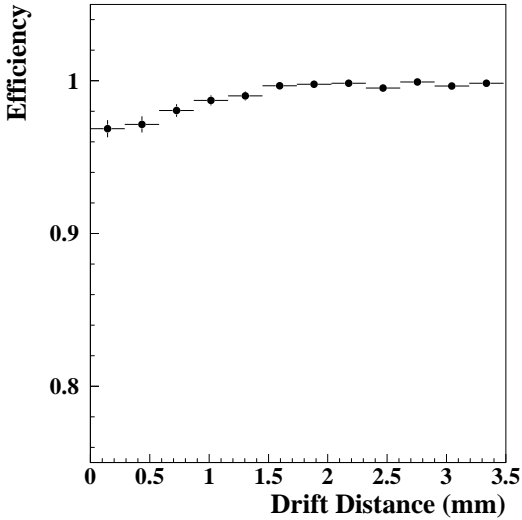,height=3.0in}
\caption{FC single-plane efficiency as function of drift distance.
\label{fig:fc-eff}}
\end{figure}

Values of the resolution and efficiency of the chambers in 1996 are
plotted as a function of track position across the drift cell in
Figs.~\ref{fig:fc-res} and \ref{fig:fc-eff}.  The values shown for
the resolution are the result of a fitting technique that removes
the contribution of the tracking resolution to the track residuals.
These values therefore better represent the resolution of the chambers.

\subsubsection{The Back Chambers.}

The back drift chambers BC1/2 and BC3/4 are in the size regime where
special attention is required in the mechanical design to maintain the
rigidity and stability necessary for uniform wire tension and precise
alignment.  These detectors are described in detail in \cite{BC-paper}.
The active areas of the modules were chosen according to
their z-positions and the acceptance of the spectrometer. The
geometrical parameters are listed in Table \ref{trtab}. Each module
contains 14 fiberglass/epoxy (GFK) frames of 8 mm thickness, including
6 wire frames. They are clamped between two 38 mm thick non-magnetic
stainless steel frames.  The total thickness of a module is 190
mm. Because of their great length, the frames were prestressed to
maintain the wire and foil tensions. The wires are located precisely
by contact with accurate pins made of Polyoxymethylene with 5~$\mu$m
tolerance, inserted in holes drilled in the GFK frames with
a precision of better than 50~$\mu$m. The holes are located along a
curved convex locus that compensates the expected stress deflection
of the frames, in order to prevent displacement of the U and V
wires. A mean deviation of 14~$\mu$m for the position of the holes was
measured. Pins precisely located in holes can be produced much more
easily than combs and allow convenient installation of the wires. The
gas is injected directly into the wire gaps and exhausted via the
window gaps. Work on the chamber gas leaks after the 1995 running 
period reduced the leak rate of the BC's to the level of diffusion
through the window foils.  The cathode wires and foils are operated at
the same negative high voltage of typically 1770~V; the anode wires
are at ground potential.  More details on the BC mechanical
construction can be found in \cite{BC-bern} \cite{BC-neun}.

The chamber BC1 is located in the fringe field of the magnet. It has
been verified by simulations using the program GARFIELD
\cite{BC-garfield} that a vertical component of 0.1~T contributes to a
deviation in the measured coordinate of $<$ 60 $\mu$m, which can be
neglected in the analysis.

The BC performance was studied extensively in test beam measurements
illuminating essentially only one drift cell per plane.  Under these
conditions the HV/threshold setting 1750~V/50~mV was found to be the
optimum working point, yielding $\sigma$ = 175~$\mu$m resolution
($\sigma= 150~\mu$m in the central region of the BC drift cell) and
96.3$\%$ efficiency at 6$\%$ cross talk.  The efficiency was
determined from the occurrence of chamber signals in a rather narrow
$\pm\,$4$\,\sigma$ corridor around the reference track, which was
defined by a silicon micro strip detector telescope
\cite{BC-nim,BC-hasch}.  \\

\begin{figure}
\hskip 0.01cm \psfig{figure=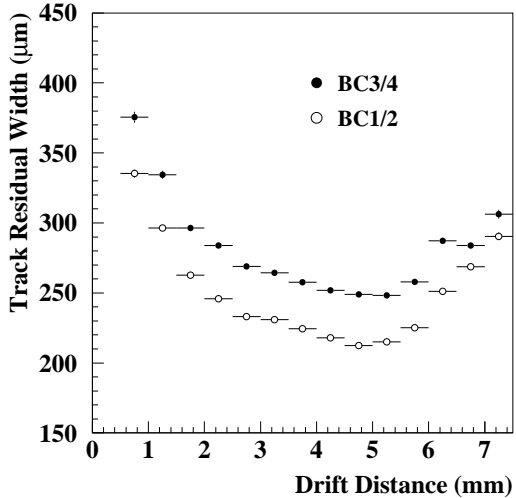,height=3.0in}
\caption{Width of the BC1/2 and BC3/4 residual distribution as 
function of the drift distance for tracks with an incident angle
$<$~1$^\circ$.  
\label{fig:bcres}}
\end{figure}

The BC performance was evaluated using production data from normal
running.  The widths of Gaussian fits to the central portions of the
residual distribution for all tracks crossing the plane with an angle
$<$~1$^{\circ}$ is given in Fig.~\ref{fig:bcres} as function of the
drift distance for both BC1/2 and BC3/4. This plot was obtained by
combining the data from all cells of all planes. Tracks close to the
sense wire (i.e. small drift distances) are excluded because this data
point is strongly affected by the difficulties in resolving the
left/right ambiguity.  However, a smooth behaviour of the residual can
be assumed. As expected, the best value is measured in the central
region of the drift cell. The results are approximately 210~$\mu$m for
BC1/2 and 250~$\mu$m for BC3/4.

\begin{figure}
\hskip 0.01cm \psfig{figure=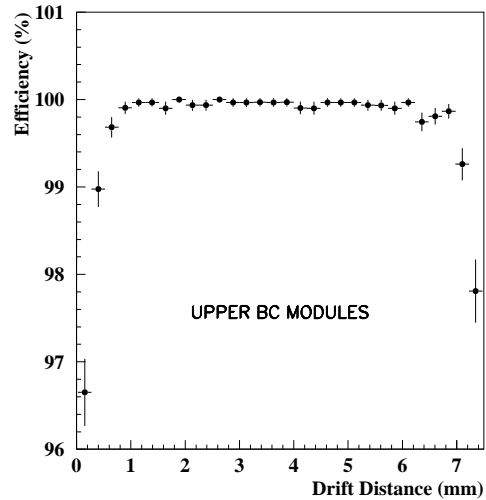,height=3.0in}
\caption{BC efficiency as a function of the drift distance for the
upper BC modules under normal running conditions.
\label{fig:bceff}}
\end{figure}

The average BC plane efficiency for electron and positron tracks is
found to be well above 99\% for 1996 data.  The situation is somewhat
worse for all tracks, which are mainly hadrons, due the smaller energy
deposited in the chambers by hadrons.  The plane efficiency drops to
97\% when all tracks are considered.  The efficiency is calculated
by using the reconstructed track as a reference and adopting
the same corridor width as used in the reconstruction program to find
the hits belonging to a track, i.e. about $\pm 3\sigma$, where
$\sigma$ is the plane resolution. The dependence of the efficiency on
the drift distance averaged over all drift cells is shown in
Fig.~\ref{fig:bceff}.

Investigations of the long term stability of the 1995 data showed the
expected correlation between chamber performance and atmospheric
pressure, which led to varying working conditions.

\begin{figure}
\hskip 0.01cm \psfig{figure=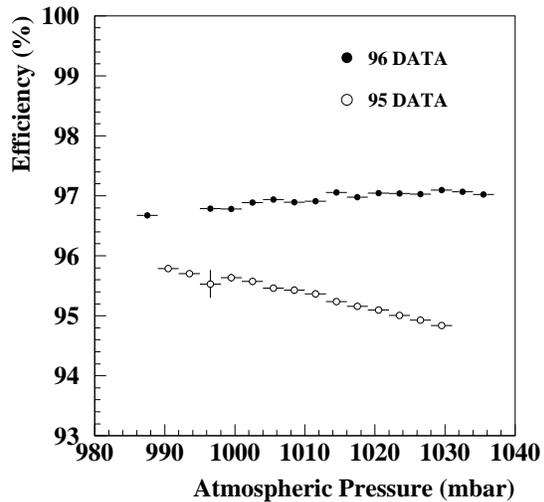,height=3.0in}
\caption{Averaged BC plane efficiency as function of the atmospheric
pressure for the 1995 data sample (open circles) and a 1996 data
sample after introducing a gain stabilisation scheme (filled
circles). For details see the text.
\label{fig:bcatm}}
\end{figure}

A system to compensate for atmospheric pressure variations by 
dynamically adjusting the high voltage was introduced during the 1996
running period.  A 'nominal' high voltage setting of 1770~V was chosen 
for an atmospheric pressure of 1013~mbar. The actual operating voltage 
was computed using a parametrisation of the gain versus atmospheric
pressure given in Ref.~\cite{BC-armit}. 
The high voltage setting is corrected in steps of 1~Volt over a range
of $\pm$~20~Volts. This scheme significantly improves the stability of
the chambers as can be seen in Fig.~\ref{fig:bcatm}.  The dependence
of the average plane efficiency on the atmospheric pressure is
presented for data before and after introduction of the gain
stabilisation scheme in the middle of 1996 running. The overall
improvement in efficiency using this system is 1 to 2 \%.  Threshold
settings of 65 mV for the BC1/2 and 80 mV for the BC3/4 have been
used.

The operation of the eight BC modules was remarkably stable and reliable
in the first 2.5 years with low dark currents.  

\subsection{The Drift Vertex Chambers}

Notwithstanding the adequate performance of the tracking system during
the 1995 commissioning year, redundancy in the front region was
insufficient. Loss of a single FC plane could have had serious
consequences. There was also foreseen the need for front tracking
elements with larger acceptance for a future upgrade to provide muon
detection beyond the standard acceptance. Hence an additional Drift
Vertex Chamber (DVC) pair was constructed, and installed immediately
following the VC's for use starting in the 1997 running period. The
DVC drift cell design is similar to but slightly smaller than that of
the FC's. However, the DVC acceptance extends vertically from
$\pm35$~mrad to $\pm270$~mrad, and covers $\pm200$~mrad horizontally.

An acute design problem arising from the DVC location near the target
chamber exit window is the close proximity of the active area to the
beam pipe. This is solved in part by separating the functions of gas
enclosure and chamber frames. The gas enclosure or box design is
similar to that of the VC's, with a very thin wall less than 1~mm from
the beam tube. Each of six wire planes in a box is clamped between its
own pair of cathode foil planes. All planes can be installed together
in the box while they are mated with the signal feed-throughs in the
lid. However, the planes are thereafter precisely aligned by two pins
equipped with gas seals, externally inserted through the box and all
the planes. The ends of these pins provide external alignment
fiducials. A gas injector is also externally inserted through the box
wall into each anode plane after its installation in the box. The
gas spills out of the active gaps into the box, and hence to a common
exit port.  Both the signal feed-throughs and the gas injectors are
designed for mechanical accommodation sufficient to avoid constraining
the alignment of the planes with respect to the box.

The readout system of the DVC's is identical to that of the FC's,
resulting in similar performance in 1997 except for one of the twelve
DVC planes, which suffered a broken wire immediately after
installation.

\subsection{The Magnet Chambers}

The proportional wire chambers MC1 through MC3 located in the gap of
the magnet were originally intended to help resolve multiple tracks in
case of high multiplicity events. Since low backgrounds have made this
unnecessary, their primary function is now the momentum analysis of
relatively low energy particles, from the decay of $\Lambda$'s for
example.  Since the MC's operate in a strong magnetic field and
resolution of $\pm$1$\,$mm is sufficient, multi-wire proportional
chambers with digital single bit-per-wire readout were chosen. The
magnet has a tapered pole configuration that implies a different size
for each pair of chambers.  The active areas of the three chambers
given in Table \ref{trtab} are dictated by the spectrometer acceptance
together with magnetic dispersion at a nominal trigger threshold
positron energy of 3.5 GeV.

Each chamber consists of three submodules $U, X$ and $V$, laminated
into one module with common gas volume. Each submodule consists of an
anode plane at ground potential and two cathode planes with a common
negative HV connection unique to this submodule, typically at
2850~V. The distance between anode and cathode planes is
4$\pm$0.03$\,$mm and the centers of neighbouring submodules are
separated by 21$\,$mm. The gas mixture has the same constituents as
the drift chamber gas, but with proportions (Ar-CO$_2$-CF$_4$ 65:30:5)
optimized for MWPC operation (see Table \ref{trtab}).

The most difficult aspect of the MC design was the limited space
inside the magnet for the frames and onboard electronics. The
electronics are mounted on the front and back faces of the chambers in
the barely adequate space between the magnet pole face and the
spectrometer acceptance. Thin flexible printed circuit foils connect
the wire frames to the electronics. The readout is the LeCroy PCOS IV
system, including an on-chamber card design configured specifically
for this application \cite{MC-LeCroy}. In addition to signal
amplification, these cards provide discrimination and delay as well as
latching in response to the event trigger. The high speed serial
readout (up to 20~Mbit/s) greatly reduces the cabling in the very
confined space of the magnet gap.  The most severe constraint on MC
operation is the inaccessibility of the detectors and the on-board
electronics for the entire running period.  It is not possible to work
on the detectors during regular monthly accesses because they are
buried inside the magnet. Unfortunately, a faulty batch of
discriminator/latch chips together with a less-than-optimal design of
the output motherboards used in 1995 created a reliability problem.
The failure of a single chip that reads out only 8 wires disabled as
many as 16 cards, corresponding to 256 wires. This process in
combination with a Low Voltage distribution malfunction eventually
resulted in the failure of a large fraction of the MC system, with
overall tracking efficiency of only 45\%. A redesigned motherboard
eliminated this problem for 1996 operation. The MC system then
performed reliably with typical efficiency per plane of 98-99\%.
Electronics failures in the 1996-97 running period resulted in the
loss of only 0.5\% of the channels (3 of 688 cards).

The very restricted air flow in combination with the power dissipation of
250\,mW per channel makes water cooling a necessity. Flexible metallic
fingers contact the tops of the SMD chips and conduct the heat to
water-cooled plates between the cards. The water system operates at a
pressure less than atmospheric to eliminate the danger of leaks.

The contribution to the spectrometer momentum resolution for 10~GeV
positrons by Coulomb multiple scattering in the magnet chambers and
air between FC2 and BC1/2 is $\theta_{rms}$ = 0.15 mrad.  (If the air
were replaced by helium, it would be decreased to 0.11 mrad.) Thus the
ratio of $\theta_{rms}$ to the magnet deflection angle is equal to
0.0024, independent of energy. This is small compared to other
contributions.

\subsection{Alignment}

Errors in the relative alignment of the various elements of the
tracking system are an important contribution to the 
resolution in reconstructed kinematic quantities. The initial
alignment during detector installation was done using conventional
optical techniques, but difficulties with optical access to the
components limited the expected accuracy of this process to a few
hundred micrometers in x and y, and of order 1~mm in z. It was always
expected that the relative detector alignment would have to be improved
in the data analysis through the study of track residuals with the
spectrometer magnet turned off. This procedure was refined over the
first three years of operation to the point that drift chamber alignment
errors are no longer a major contribution to tracking resolution
\cite{Align-Gaerber}.

\subsubsection{Laser Alignment System}

It was considered essential to have a means to continuously monitor
the relative alignment of the tracking detectors for two reasons.
First, the detectors are mounted on a complex support system that is
not simply interconnected so that its response to temperature changes
and gradients is difficult to predict. Also, the front detectors are
supported ultimately via the spectrometer magnet field clamp, which
might suffer distortion from magnetic forces. Such an effect was
detected with this monitoring system, and eliminated by stabilizing the
clamp mount.  Secondly, if a detector must be replaced with a spare
after the laborious alignment analysis has been done, it is
desirable to be able to avoid having to repeat this process
immediately with new field-off data.

\begin{figure}
\hskip 2.0cm \psfig{figure=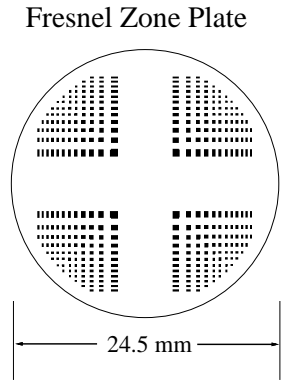,height=2.0in}
\caption{Fresnel zone plate made of 13 $\mu$m thick
stainless steel with a diameter of 24.5 mm.
\label{fig:fresnel}}
\end{figure}

A solution to both of these problems is to have on each detector module
optical targets that are continuously accessible to a remotely
operated image recording system \cite{Align-slac} . The type of system
that was chosen employs two well-collimated parallel laser beams
passing through the entire spectrometer in the beam plane, on either
side of and parallel to the beam axis. The 2~cm diameter beams are
created by one He-Ne laser and a semi-transparent mirror. Two optical
target assemblies are mounted precisely and reproducibly on each
detector element. Each assembly includes a remotely-controlled
actuator that can move an optical target into the path of a laser
beam, arriving at a precise and reproducible stop. Only one target is
in a laser beam at a time.  The type of target used -- the Fresnel
zone plate -- is similar to that used in the SLAC accelerator
alignment monitoring system.  It is made by etching a precise pattern
through a thin metal foil. This avoids any deflection of the beam from
imperfections in the planarity of transparent material.  A Fresnel
zone plate is shown in Fig.~\ref{fig:fresnel}.  The large unperforated
central regions were left for the sake of mechanical stability.  They
have the disadvantage that they produce secondary fringes in the
images.

The zone plates on each detector are specified with a focal length
equal to the optical path length from the target to the image
recording component, which is a lensless CCD camera with a pixel size
of 11 ${\mu}$m ${\times}$ 11 ${\mu}$m. Hence any shift in the detector
or target is reproduced with unity magnification as a shift in the
focal pattern at the camera. The camera image frames are digitized and
recorded via a VME based system.

Since there are two targets on each detector at x=$\pm$45~cm, all
transverse detector coordinates can be monitored. No absolute
alignment is attempted with this system. In fact, long term stability
of the laser beams is not assumed, since the optical paths contain
several mirrors that are subject to instability. The beams must remain
stable only for the 10 minute period required to cycle through all
targets in each beam path. Only the relative x and y positions of the
images recorded within that cycle are interpreted as being
significant. Thermal wavering of the beams is reduced by insulating
the beam path through the warm magnet shielding plate using a
foam-cell tube. Several camera frames are averaged to minimize the
effects of the wavering.

\begin{figure}
\hskip 1.0cm \psfig{figure=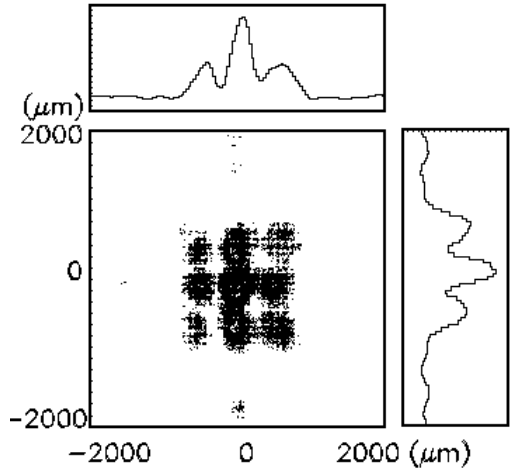,height=2.5in}
\caption{Typical Fresnel pattern image detected by the CCD camera.
A one-dimensional projection is shown at the top and on the right.
Nine peaks are visible at the center.  The position of the
central peak is fitted.
\label{fig:ccd}}
\end{figure}

A Fresnel pattern image is shown in Fig.~\ref{fig:ccd} .
The FWHM of the central peak is about 300 ${\mu}$m.
The precision of the laser alignment system is 30 - 50 ${\mu}$m.

\subsection{Track Reconstruction}

The \h\ reconstruction program (\hrc) \cite{HRC-hrc} is very efficient
because it makes use of two unusual methods: the tree-search algorithm
for fast track finding and a look-up table for fast momentum
determination of the tracks.  The full reconstruction of 30 Monte
Carlo events on an SGI R4400 processor takes only one second.

\subsubsection{The Fast Pattern Recognition Algorithm}

The main task of the reconstruction program is to identify particle
tracks using the hits in the tracking chambers.  Each detector plane
gives spatial information in one coordinate, and only by combining the
information of many detectors is it possible to reconstruct the tracks
uniquely in space. There are several track-finding algorithms in
common use. \hrc\ uses the tree-search algorithm, which turns out to be
very fast.  This algorithm is implemented in the following way:

As a first step, ``partial'' tracks have to be found in projections,
separately in the region in front of and behind the spectrometer
magnet. The track projections are approximately straight lines in
those regions, except for small curvatures caused by the magnetic
fringe fields, and kinks coming from secondary interactions and
straggling.

\begin{figure}
\hskip 1.0cm \psfig{figure=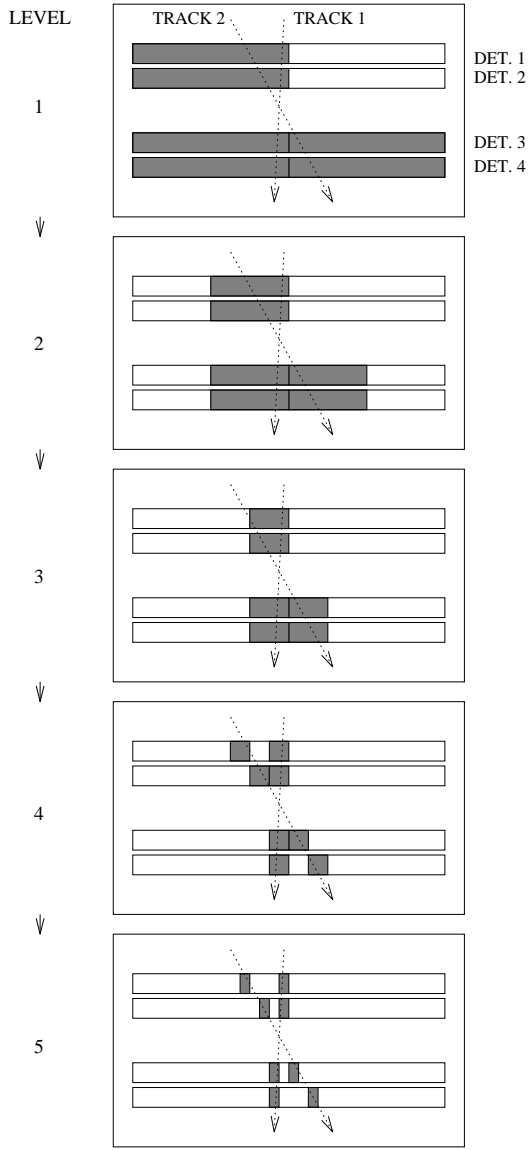,height=6in}
\caption{The tree search algorithm looks at the hits of the
tracking detectors with artificially reduced resolution. In every tree
search level the resolution is doubled until a resolution is reached
that is optimal for track finding.
\label{fig:treesearch}}
\end{figure}

The basic idea of pattern recognition using the tree-search algorithm
is to look at the whole hit pattern of the detectors with variable
(increasing) resolution as illustrated in Fig.~\ref{fig:treesearch}.
At each step of the iteration the
entire partial track is seen.  However only at the end is the full
detector resolution reached. The detector resolution is roughly
250~$\mu$m and the active width of a chamber is of the order of a few
meters. Therefore after about 14 steps in the binary tree, the
resolution of the detector of $\sim 1:2^{14}$ would be reached. For
the purpose of track finding (not fitting) a resolution of $1:2^{11}$
is sufficient, which reduces the maximum number of iterations in the
\hrc\ tree-search to about 11.

\begin{figure}
\hskip 1.0cm \psfig{figure=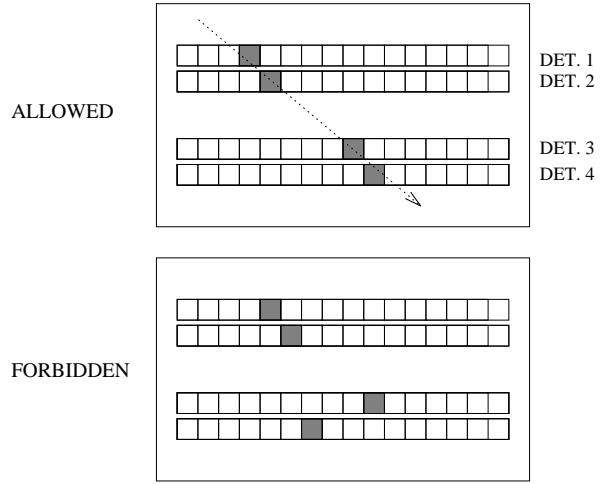,height=2.5in}
\caption{  At each tree search level, the allowed and forbidden
patterns are well defined. The program compares the measured pattern
with the data base and determines if the measured pattern is consistent
with an allowed track.
\label{fig:treeallowed}}
\end{figure}

In each step of the iteration, the algorithm checks if the pattern (at
the given resolution) contains a sub-pattern that is consistent with
an allowed track. Fig.~\ref{fig:treeallowed} shows examples of allowed 
and forbidden patterns. All allowed patterns are generated and
stored in a data base at the initialisation phase of the program. The
comparison is very fast as only look-up tables are used; no
calculations have to be done during event processing.

The number of allowed track patterns at the full resolution is of the
order of 10$^8$. Without optimization, this large number would make the
algorithm useless for two reasons: memory space problems
to store all the patterns, and CPU time problems to compare with each
of them. The following methods avoid these problems: 
\begin{itemize}
\item  If one compares  the allowed patterns from one iteration to
those from the following one (called parent and child in the following)
it becomes obvious that only a very limited number of children exist for
each parent.  At initialisation time all links between the allowed
patterns of each generation are calculated, i.e.~all children for each
parent are stored in the data base. The number of comparisons now becomes
small: if a pattern is recognised at one generation, only its children 
have to be checked in the next generation.  The number of children
for each parent is typically 4 to 8. This solves the CPU problem.
\item Symmetry considerations can be used to reduce the number of patterns
that have to be stored. If two patterns are mirror symmetric, or if
they are identical except for a transverse shift, they are stored as
one pattern.  Most important is that if a child is identical to a
parent, then the child is linked to the parent and all grand children
become identical to the children. This simplification is
based on the fact that we are looking only for approximately straight
tracks. This reduces the number of
branches in the tree-search significantly.  The number of 
patterns in the data base is of the order of 50,000, which is small
compared to the original number of 10$^8$.
\end{itemize}
After applying the tree search algorithm to the $U$ and $V$ planes,
the partial tracks in these projections are defined. They are called
tree-lines.  By testing all combinations of tree-lines and merging
them with hits in the $x$ coordinate, the partial tracks in space are
found. The tree-search algorithm is not applied to the $x$-coordinate
directly as portions of
the $X$-planes of the VC chambers are tilted and thus do
not fulfil the symmetry conditions. However, the $x$ projections 
are used for track finding in the back region.

The procedure described until now is applied independently to the front
and back regions, resulting in a set of front partial tracks and
another set of back partial tracks. All combinations of front
and back partial tracks are tested to see if they match spatially
within a specified tolerance at the x-y plane in the center of the
magnet. Those combinations that match are combined to form a
full track, after refitting the track geometry to the chamber hits,
subject to the match condition. The matching condition is refined by
various small corrections depending on partial track geometry and
deflection angle from the first iteration. 

\subsubsection{The Fast Momentum Look-Up}

As a second feature in \hrc , a very fast method has been developed to
determine the momentum of a track, which is given by the deflection in
the inhomogeneous field of the spectrometer magnet. Tracking through a
magnetic field is very CPU intensive. The new method makes obsolete
the tracking through the magnetic field on a track by track basis.
Instead, a large look-up table is generated only once during
initilization.  It contains the momentum of a given track as a
function of the track parameters in front of and behind the
magnet. The relevant track parameters are the position and the slope
of the track in front of the magnet and the horizontal slope behind
the magnet. The resolution of the table has been chosen such that,
using interpolation methods, the contribution by \hrc\ to the
precision of the track momentum determination is better than $\Delta
p/p=0.5\%$. The look-up table contains 520,000 numbers.  This method
of momentum determination is extremely efficient.

The vertical slope and position of the track behind the magnet are
not kinematically relevant degrees of freedom and are used only to
determine the track quality and reduce the number of ghost
tracks. They are not used in the determination of the kinematic
parameters, as the resolution and alignment of the VC-FC system is
superior to the resolution of the back chambers.

\subsubsection{Tracking in 1995 without VC's and DVC's}

As mentioned earlier, the VC performance in 1995 was not optimal due
to difficulties in production of the APC readout chips.  The
performance of the substitute prototype chips resulted in marginal
efficiencies (60-90\%) and many "hot" channels. The DVC's were
conceived later and became available only for 1997.  Hence an
alternative track reconstruction method was developed using only the
drift chamber data. The back partial tracks are generated as usual and
the front partial tracks are generated using only the FC data. The
front-back matching search is done as usual, except with somewhat
larger tolerances. Then, for all matching pairs of partial tracks, the
match point at the centre of the magnet defined by the back partial
track is used to refine the front partial track by pivoting it about a
conserved space point midway between FC1 and FC2. Thus the front
partial track is forced to agree at the magnet midpoint with the
presumably higher quality information from the back partial
track. This process was used successfully for the 1995/96 physics
analysis, with approximately a factor of two loss of resolution in
kinematic quantities relative to what could be expected if the VC's
were fully operational.

\subsection{Tracking System Performance}

\begin{figure}
\hskip 0.01cm \psfig{figure=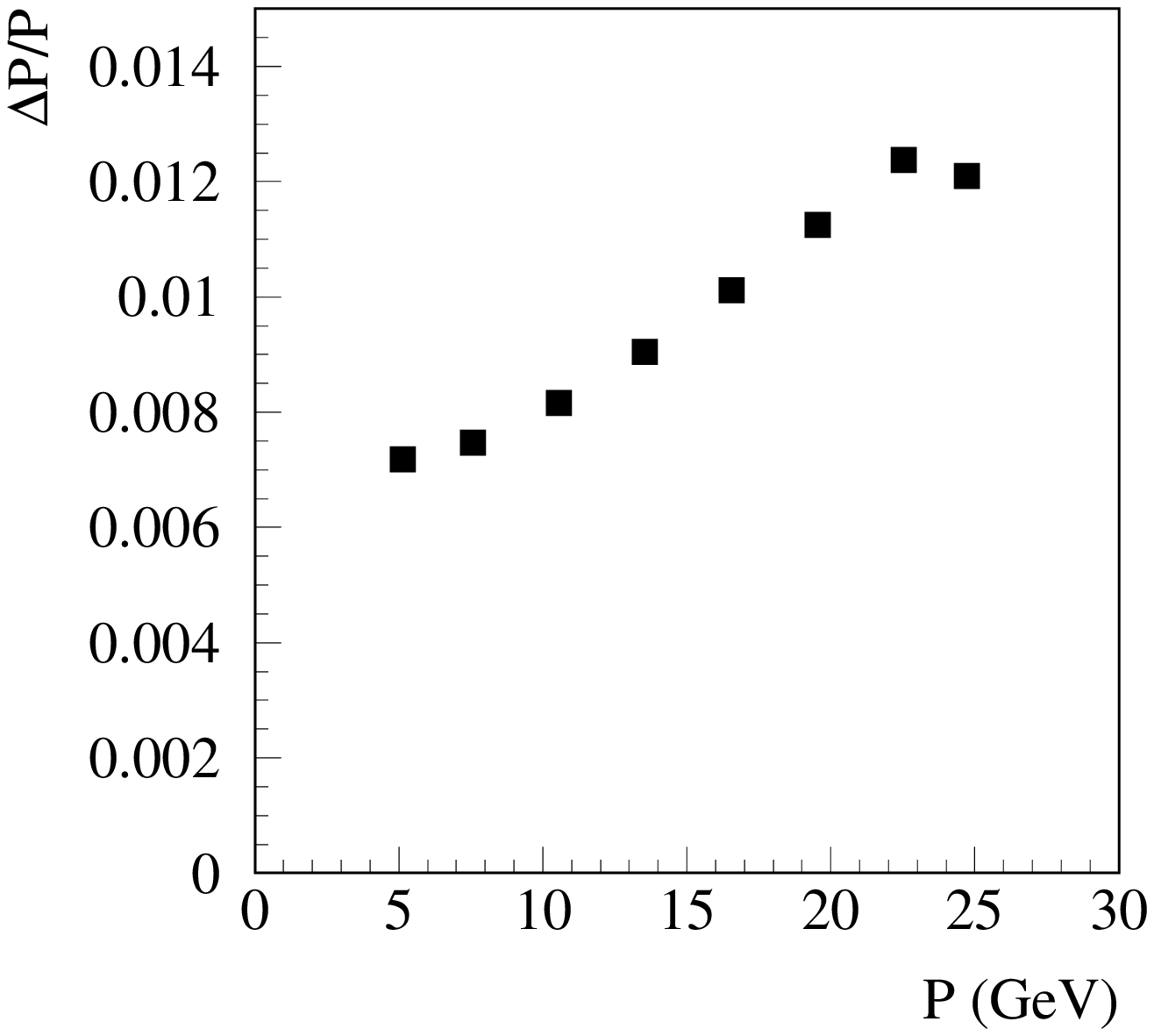,width=8.0cm}
\caption{Momentum resolution in the HERMES spectrometer, deduced
from Monte Carlo studies.
\label{fig:presol}}
\hskip 0.01cm \psfig{figure=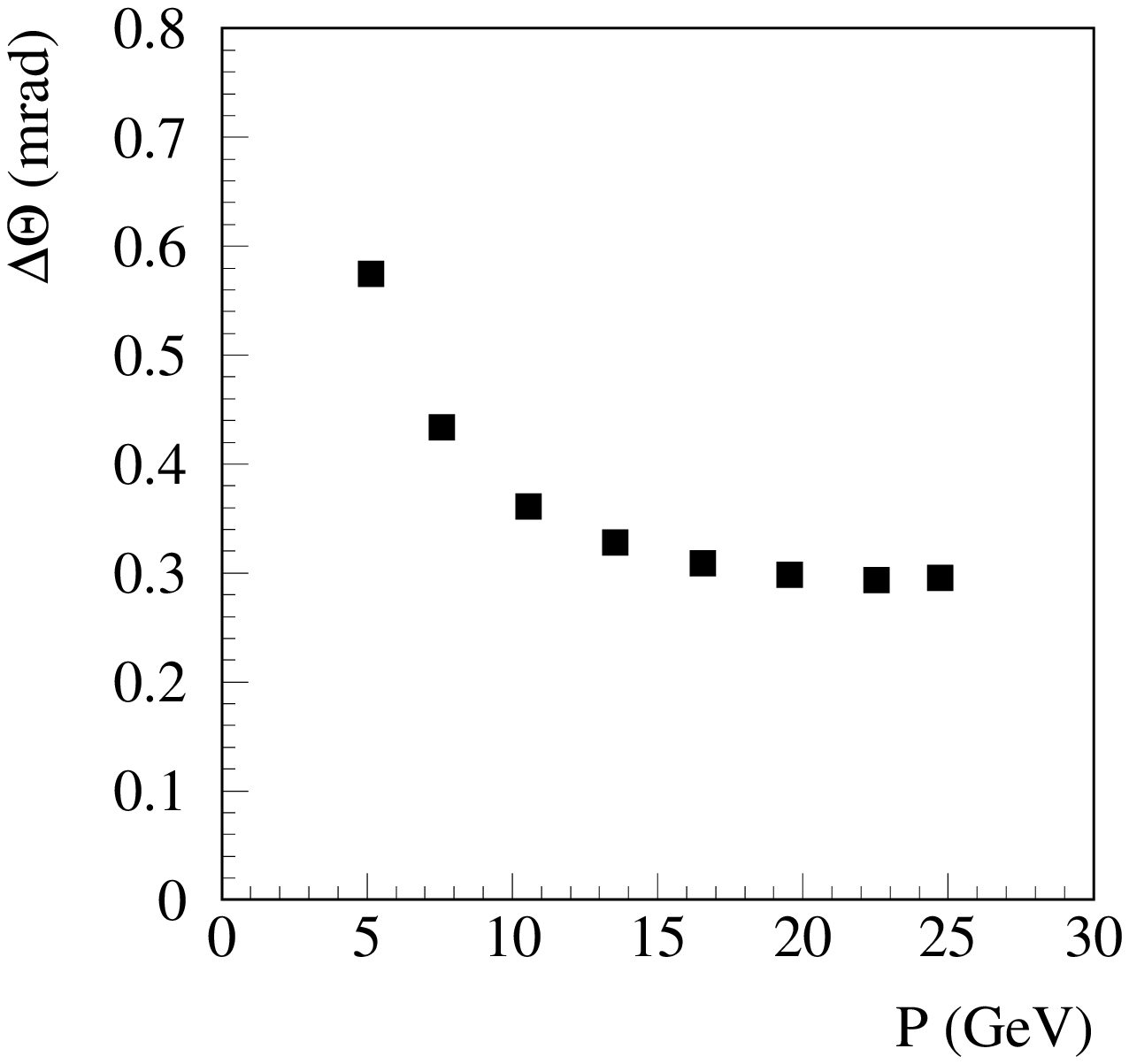,width=8.0cm}
\caption{Resolution in scattering angle deduced from Monte Carlo studies.
\label{fig:thresol}}
\end{figure}

The performance of the tracking system has been studied using the
detailed HERMES Monte Carlo (HMC) simulation of the experiment, which
is based on the GEANT \cite{GEANT} software package.  Resolutions for
several quantities were estimated by performing a Gaussian fit to
spectra of the difference between the generated and reconstructed
quantities.  Results for the measured quantities p (momentum) and
$\theta$ (scattering angle) are shown in Figs.~\ref{fig:presol},
\ref{fig:thresol}.  The momentum resolution is 0.7-1.25\% over the
kinematic range of the experiment, while the uncertainty in the
scattering angle is below 0.6~mrad everywhere.  Bremsstrahlung in
materials in the positron path cause the momentum resolution function
to have a standard deviation considerably larger than that of the
fitted Gaussian.
\begin{figure}
\hskip 0.01cm \psfig{figure=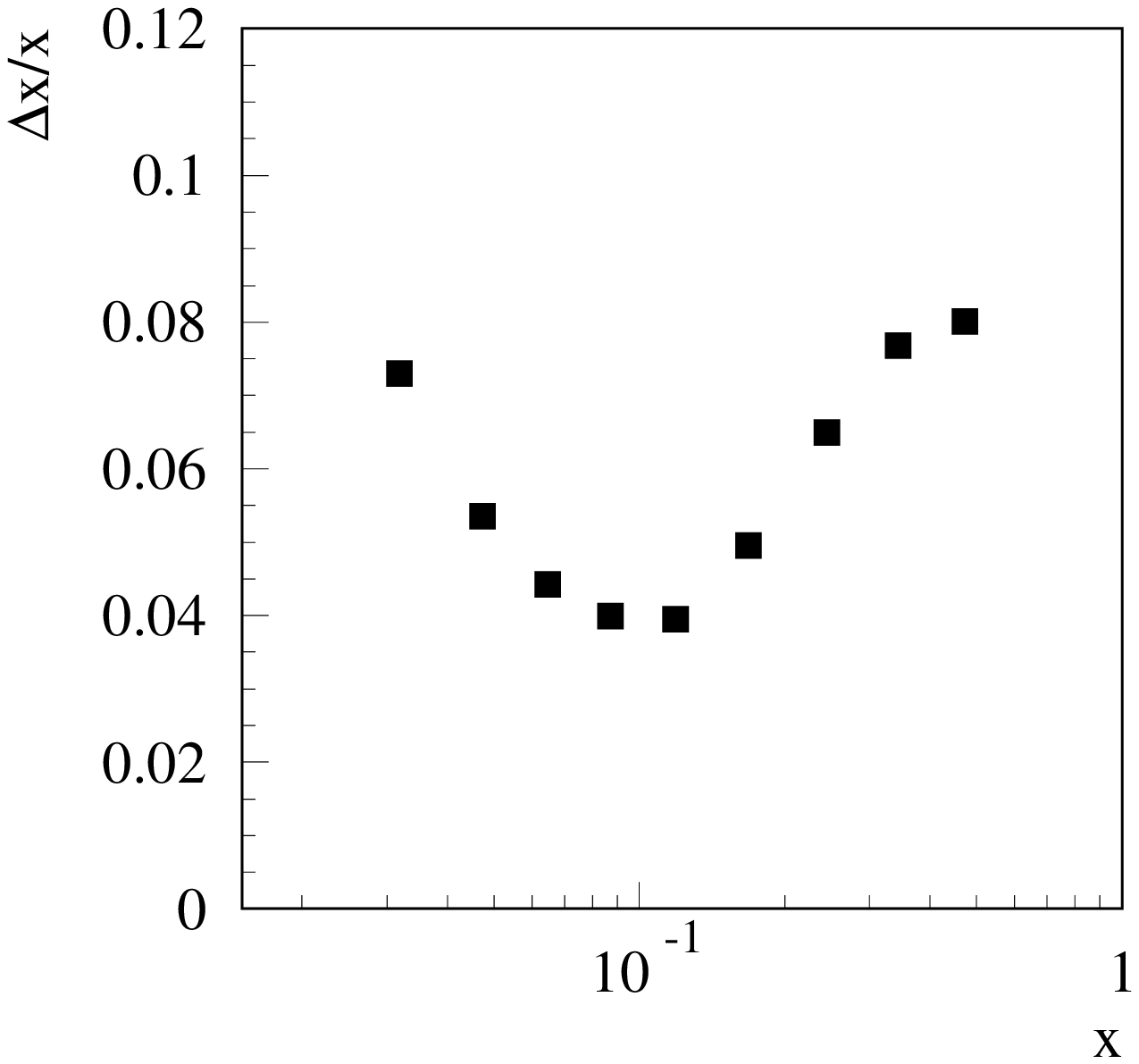,width=8.0cm}
\caption{Resolution for the Bjorken $x$ variable, deduced from Monte
Carlo studies.
\label{fig:xresol}}
\hskip 0.01cm \psfig{figure=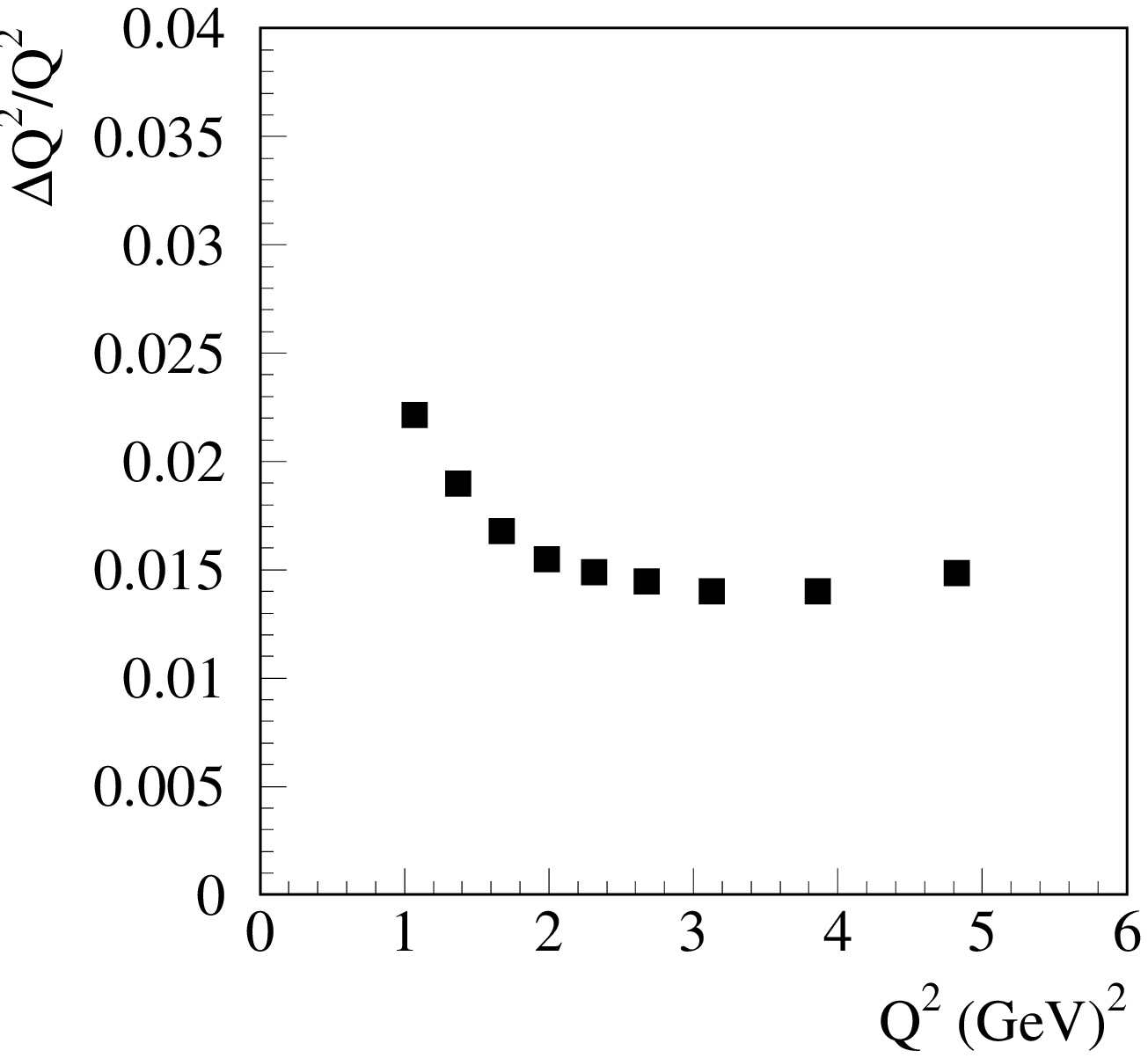,width=8.0cm}
\caption{Resolution for the square of the four-momentum transfer, Q$^2$,
deduced from Monte Carlo studies.
\label{fig:q2resol}}
\end{figure}
The resolutions for $x$ and the square of the momentum
transfer are shown in Figs.~\ref{fig:xresol}, \ref{fig:q2resol}.  The
$x$ resolution varies from 4\% to 8\% while the Q$^2$ resolution is
better than 2\% over most of the kinematic range.

\begin{figure}
\hskip 0.1cm \psfig{figure=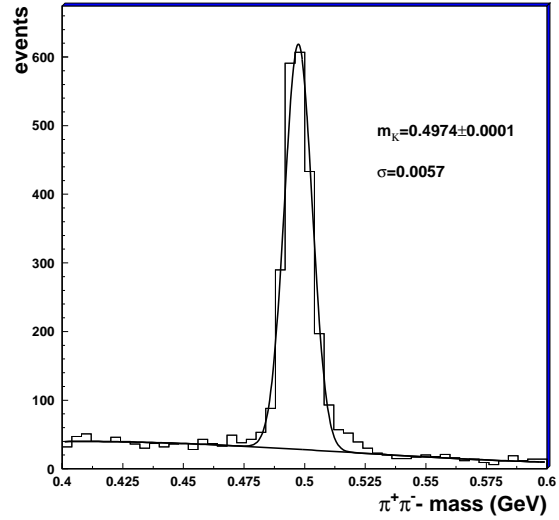,width=7.0cm}
\caption{Invariant mass of $\pi^+-\pi^-$ pairs.  The reconstructed K$_S$
mass agrees very well with the PDG value.
\label{fig:kshort}}
\end{figure}

Finally, the absolute calibration of the spectrometer can be checked
using the decay of K$_S$ mesons into two pions.  The invariant
two-pion mass is plotted in Fig.~\ref{fig:kshort}.  The reconstructed
K$_S$ mass agrees to one part per mil with the value of the Particle
Data Group \cite{PDG} (497.4~MeV vs 497.7~MeV).  The width of the peak 
($\sigma$=~0.0057, or $\approx$1.1\%) agrees with the resolution
determined by Monte Carlo methods.

\subsection{Kinematic Variables}

\begin{figure}
\hskip 0.01cm \psfig{figure=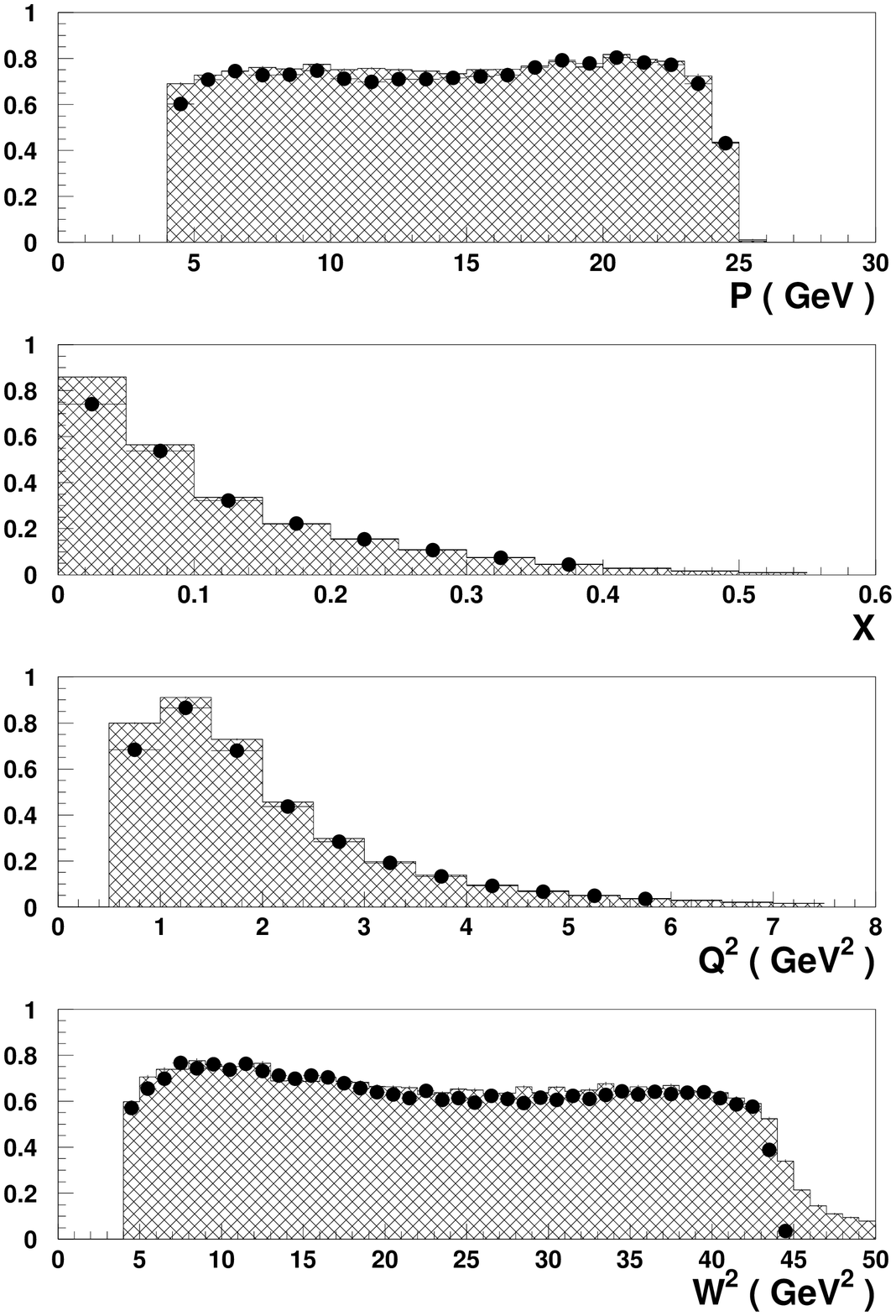,width=8.0cm}
\caption{Kinematic variables p, $x$, Q$^2$, and W$^2$.  The points are 
data while the shaded histogram is a Monte Carlo simulation.  The 
normalisations of the histograms are arbitrary.
\label{fig:kin}}
\end{figure}

Numerous measured and calculated quantities have been compared to the
Monte Carlo simulation of the experiment (HMC).  Fig.~\ref{fig:kin}
shows some of these comparisons for the momentum p of the scattered
particles, the variable $x$ defined earlier, the four-momentum
transfer squared (Q$^2$), and the square of the photon-nucleon
invariant mass (W$^2$).  The normalisations of the histograms are
arbitrary. The points represent data while the shaded histograms are
the Monte Carlo simulation.  The agreement is good over most of the
range of the variables.

\begin{figure}
\hskip 0.01cm \psfig{figure=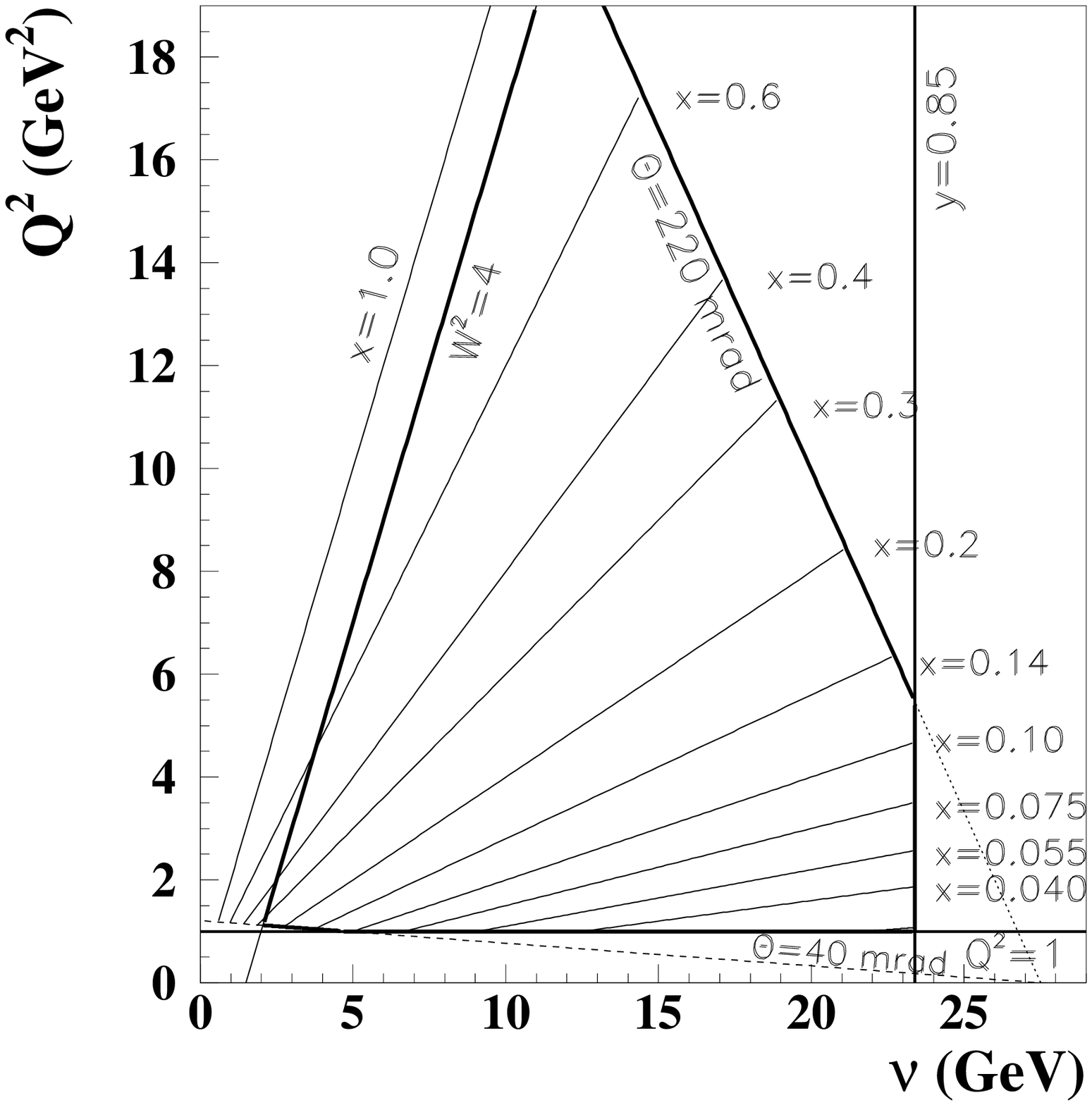,width=8.0cm}
\caption{Kinematic plane for the HERMES experiment.  See text for details.
\label{fig:kinplane}}
\end{figure}

The kinematic plane for the experiment (Q$^2$ vs $\nu$) is shown in
Fig.~\ref{fig:kinplane}.  The typical DIS acceptance is defined by the
following software cuts shown as the bold lines on the plot:

\noindent
- Scattering angle: 40~mrad $< \theta <$ 220~mrad;\\
- Fractional energy transfer: y= $\nu$/E $<$ 0.85;\\
- W$^2>4$ GeV$^2$;\\
- Q$^2>1$ GeV$^2$.

\section{The Particle Identification Detectors}

\subsection{General Description}

The HERMES particle identification system discriminates between
positrons, pions, and other hadrons.  It provides a factor of at least
ten in hadron suppression at the DIS trigger level to keep data
acquisition rates reasonable.  The rate of DIS positrons is much
exceeded by that of hadrons from photoproduction by a factor as high
as 400:1 in certain kinematic regions.  The system provides a hadron
rejection factor (HRF) of at least 10$^4$ in offline analysis to keep
the contamination of the positron sample by hadrons below 1\% for the
whole kinematic range.  The HRF is defined as the total number of
hadrons in the spectrometer acceptance divided by the number of
hadrons misidentified as positrons.  It depends on the selected
efficiency of positron identification (see Table \ref{pidtable} later
for typical efficiencies and contaminations).  Since most of the
hadrons are pions, we will often refer to a pion rejection factor
(PRF).  The PID system also discriminates pions from other hadrons for
the important semi-inclusive measurements that will allow the
contribution of the valence and sea quarks to the nucleon spin to be
isolated.

The PID system consists of four sub-systems: a lead-glass calorimeter,
two plastic scintillator hodoscopes, one of which is preceded by two
radiation lengths of lead and which acts as a pre-shower detector, a
transition radiation detector, and a threshold \v{C}erenkov detector
(see Fig.~\ref{fig:spect}).  The calorimeter and the hodoscopes are
used in the first level trigger to select DIS events.  Beam tests at
CERN have shown that this combination gives a HRF of several thousand
for an electron efficiency of 95\% in offline analysis.  The rejection
factor is estimated to be approximately 10 in the trigger.  The TRD
consists of 6 modules in each half and provides an additional HRF of
over 100 for 90\% e$^+$ efficiency (several hundred if a probability
based analysis is used).  The main function of the threshold
\v{C}erenkov detector is to distinguish pions from other hadrons for
the semi-inclusive measurements.  In future, kaons will also be
identified using a ring imaging \v{C}erenkov detector (RICH).  An
upgrade of the current threshold detector to a RICH is being prepared
for 1998.  Each sub-system is described in detail in the following
sections.

\subsection{The Calorimeter}

The function of the calorimeter is to provide a first level trigger
for scattered positrons, based on energy deposition in a localized
spatial region ($\geq$3.5 GeV in 1995 and the beginning of 1996 and
$\geq$1.4 GeV in late 1996 and 1997); to suppress pions by a factor of
$\geq$10 at the first level trigger and $\geq$100 in the off-line
analysis; to measure the energy of positrons and also of photons from
radiative processes or from $\pi^0$ and $\eta$ decays.

The solution chosen to meet these requirements consists of radiation
resistant F101 lead-glass blocks \cite{CAL-f101} arranged in two walls
of 420 blocks each above and below the beam.  The properties of F101
glass are listed in Table \ref{tabf101}.

\begin{table}[hbt]
\centering
\begin{tabular}[t]{|l|r|} \hline
\underline{Chemical composition (weight \%)} & \\
$\rm Pb_3O_4$  &  51.2 \\ 
$\rm SiO_2$    &  41.5 \\ 
$\rm K_2O$     &   7.0 \\ 
Ce             &   0.2 \\ \hline
Radiation Length (cm) &  2.78 \\ \hline
Density (g/cm$^3$)    &  3.86 \\ \hline
Critical Energy (MeV) & 17.97 \\ \hline
Moli\`ere radius (cm) &  3.28 \\ \hline
Refractive Index      &  1.65 \\ \hline
Thermal Expansion coefficient (C$^{-1}$) & 8.5 x 10$^{-6}$ \\ \hline
\end{tabular}
\caption{Properties of F101 Glass. 
\label{tabf101}}
\end{table}

Each block is viewed from the rear by a photomultiplier tube (PMT).
The blocks have an area of 9$\times$9~cm$^{2}$, a length of 50 cm
(about 18 radiation lengths), and are stacked in a 42$\times$10 array.
The blocks were polished, wrapped with 0.051~mm thick aluminized mylar
foil, and covered with a 0.127~mm thick tedlar foil to provide light
isolation.  Each block is coupled to a 7.62~cm Philips XP3461 PMT by a
silicone glue (SILGARD 184) with refractive index 1.41.  A $\mu$-metal
magnetic shield surrounds the PMT. A surrounding aluminum tube houses
the $\mu$-metal and provides the light seal.  It is fixed to a flange
that is glued to the surface of the lead glass.  This flange is made
of titanium, matching the thermal expansion coefficient of F101.

To prevent radiation damage of the lead glass, both calorimeter walls
are moved away vertically from the beam pipe by 50~cm for beam
injection.  The monitoring of gain and ageing is achieved using a dye
laser at 500~nm, which sends light pulses of various intensities
through glass fibres to every PMT of the calorimeter, and additionally
to a reference counter photodiode.  The various intensities are
produced by a rotating wheel with several attenuation plates. The
light is split in several stages before being fed into the glass
fibres. As the gain of the photodiode is stable, the
ratio of the PMT amplitude to that of the photodiode signal can be
used to monitor relative gain changes in the photomultipliers.
Over three years of operation, there has been no observed degradation
of performance that would suggest ageing effects.

Radiation damage to the lead-glass is also monitored indirectly using
TF1 \cite{CAL-tf1} blocks placed behind the calorimeter.  This
material is much more sensitive to radiation damage than F101.
Therefore, a degradation of the response would be seen much sooner in
these monitor detectors if there were a large radiation dose incident
on the back of the calorimeter caused by showers produced by beam loss
in the proton machine.  So far no variation has been observed in their
response, indicating that the effect of radiation damage is negligible.

Measurements with 1-30 GeV electron beams were performed at CERN and
DESY with a 3$\times$3 array of counters. Each lead-glass block was
also calibrated within $\approx$1$\%$ at DESY in a 3 GeV electron
beam incident at the center of the block.

The performance of a 3$\times$3 array of counters showed
\cite{CAL-AV}: {\it i}) an energy response to electrons linear within
1\%, over the energy range 1-30 GeV; {\it ii}) an energy resolution
that can be parameterized as $\sigma (E) /E \; [\%] = (5.1 \pm 1.1)/
\sqrt{E \, [GeV]} + (1.5 \pm 0.5)$: this is similar to that obtained
for other large lead-glass calorimeters; {\it iii}) a spatial
resolution of the impact point of about 0.7 cm ($\sigma$); {\it iv}) a PRF
of $\approx (2.5 \pm 1.2) \cdot 10^{3}$ integrated over all
energies in combination with the preshower detector, for a 95$\%$
electron detection efficiency.

\begin{figure}
   \psfig{figure=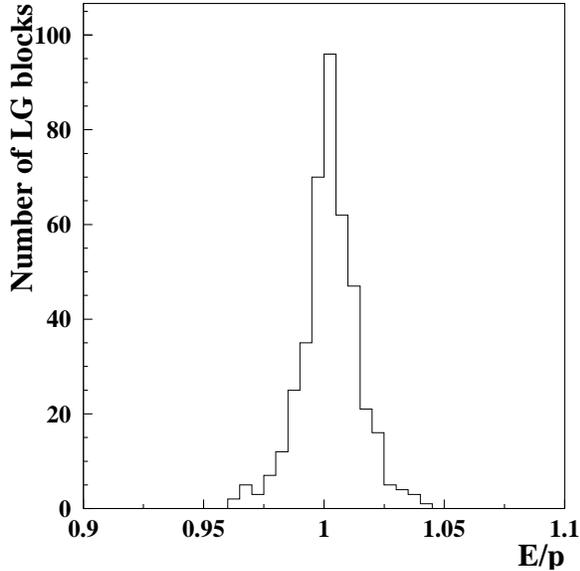,width=8cm}
   \caption{Central values of the $E/p$ distribution for each block 
     (one entry each) measured for positrons for data collected during 
     the first year of operation.
   \label{fig:ep}}
 \end{figure}

The central values of $E/p$ distributions for positrons, measured for
tracks incident on each block during the 1995 data taking period are
plotted in Fig.~\ref{fig:ep}.  $E$ and $p$ are respectively the energy
of positrons measured by the calorimeter and the momentum determined
by the spectrometer. The ratio has a central value of 1.00 with a width
($\sigma$) of 0.01, demonstrating that the response of the counters
is uniform to $\approx$ 1\%.
The long term stability of the detector response of 1$\%$ is determined
from the mean value of the $E/p$ distribution, measured for each run
for the counters nearest to the beam.  This value also includes the
contribution of the radiation damage produced in one year of operation,
indicating the effect of the radiation damage is negligible.

\begin{figure}
   \psfig{figure=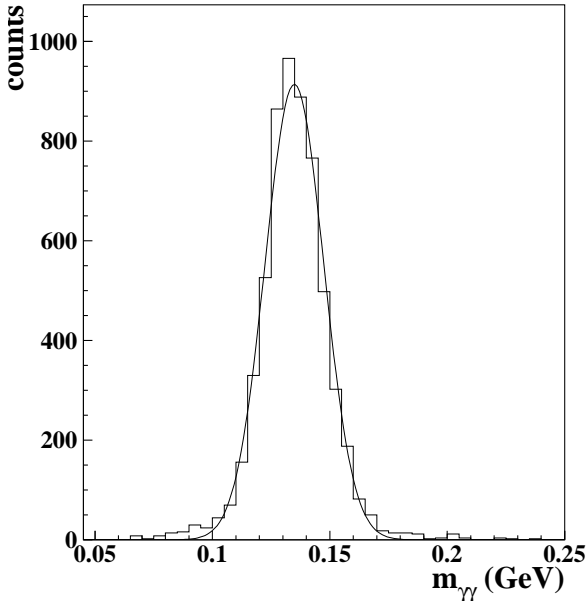,width=8cm}
   \caption{$\pi^0$ mass reconstructed from 2 $\gamma$ cluster events.
   \label{fig:pi0}}
 \end{figure}

The overall calibration of the calorimeter can be determined by the
reconstruction of $\pi^0$ decays.  Fig.~\ref{fig:pi0} shows the
$\pi^0$ invariant mass for events with two $\gamma$ clusters.  The
centroid of the peak for the 1995 data is (134.9$\pm$0.2) MeV with a
$\sigma$= (12.5$\pm$0.2) MeV in good agreement with the particle data
group value \cite{PDG}.  The calorimeter is described in more 
detail in a separate paper \cite{CAL-paper}.

\vspace{1cm}

\subsection{The Hodoscopes}

A scintillator hodoscope (H1) and Pb-scintillator preshower counter
(H2) provide trigger signals and particle identification information.
Both counters are composed of vertical scintillator modules (42 each
in the upper and lower detectors), which are 1~cm thick and 9.3~cm x
91~cm in area.  The material for the modules is BC-412 from Bicron Co.,
a fast scintillator with large attenuation length (300 - 400 cm). The
scintillation light is detected by 5.2~cm diameter Thorn EMI 9954
photomultiplier tubes, one coupled via a light guide to the outside
end of each scintillator (away from the beam plane). The modules are
staggered to provide maximum efficiency with 2 - 3~mm of overlap
between each unit.

\begin{figure}
\hskip 0.01cm \psfig{figure=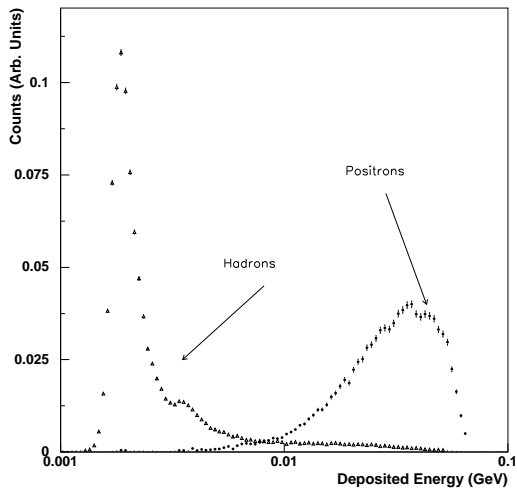,height=3.0in}
\caption{Response of the pre-shower detector to hadrons and positrons.
\label{fig:hodo}}
\end{figure}

In addition to providing a fast signal that is combined with the
calorimeter and H1 to form the first level trigger (H0 was implemented
in 1996; see below), the H2 counter provides important discrimination
between positrons and hadrons. This is accomplished with a passive
radiator that initiates electromagnetic showers that deposit typically
much more energy in the scintillator than minimum ionizing particles.
The passive radiator consists of 11 mm (2 radiations lengths) of Pb,
sandwiched between two 1.3~mm stainless steel sheets. The response of
the H2 counter to positrons and hadrons is shown in Fig.~\ref{fig:hodo}.
While pions deposit only about 2 MeV, positrons produce a broad
distribution of deposited energies with a mean of 20 - 40 MeV that
depends weakly on the energy of the incident positron. A PRF of
$\sim$~10 is possible with 95\% efficiency for positron detection.

\subsection
{Forward Trigger Scintillators}

In 1995, a large number of showers generated by the proton beam were
able to satisfy the trigger requirements for DIS positrons.  There
were typically between 20 and 100 triggers per second of this type,
compared to about 35 positron triggers and 35 pion-induced triggers
per second (with the standard $^3$He target operation and about 30 mA
of positron current).

A forward trigger scintillator (H0) placed directly upstream of the
front drift chambers was introduced in 1996 to eliminate these
triggers by distinguishing forward and backward going particles using
the time of flight between the front and rear scintillators.

The total time required for a light-speed particle to traverse all
trigger detectors is on the order of 18~ns, and hence a backward-going
particle produces a pulse in the front counter that is displaced in
time by about 36 ns from the normal trigger condition.  Such a large
time displacement allows for easy elimination of these background
events in the trigger.

The front trigger detectors consist of a single sheet of standard
plastic scintillator, 3.2 mm thick (0.7\% of a radiation length).  
The occupancy of the scintillators is such that segmentation of the
detector is not necessary.  The rate in each paddle is about $10^6$
per second.  The scintillation light is collected along the edge
farthest from the beam axis into two sets of lucite strips and then
into two 5.08~cm Thorn EMI 9954SB phototubes.

\subsection{The \v{C}erenkov Detector}

Pion identification is provided by a pair of single-gas radiator
threshold \v{C}erenkov counters, one each above and below the beam.
The units are located between the back drift chamber groups, BC1/2 and
BC3/4.  The radiator is a gas mixture at atmospheric pressure of
nitrogen and perfluorobutane, ${C_4}{F_{10}}$, the composition of
which can be varied to control the momentum interval over which pions
can be distinguished unambiguously from other hadrons.  The depth of
the radiator volume is 1.17~m.  During the first year of operation,
the radiator was pure nitrogen for which the \v{C}erenkov momentum
thresholds for pions, kaons, and protons are 5.6, 19.8, and 37.6~GeV,
respectively.  In 1996-97, a mixture of 70\% nitrogen and 30\%
perfluorobutane was used, which gives corresponding thresholds of 3.8,
13.6, and 25.8~GeV.

The body of each unit is constructed of aluminum.  The entrance and
exit windows have areas 0.46~m $\times$ 1.88~m and 0.59~m $\times$ 2.57~m,
respectively.  Each window consists of a composite foil of 100 $\mu$m
of mylar and 30~$\mu$m of tedlar separated by a 1~cm gas gap from a
second identical composite foil.  Dry nitrogen gas continuously
flowing through this gap in each window provides a barrier against the
diffusion of atmospheric gas through the windows into the radiator gas
volume.  An array of 20 spherical mirrors (radius of curvature:
156~cm) mounted at the rear of the counter on Rohacell forms focuses
the \v{C}erenkov light emitted along particle trajectories onto
corresponding members of an array of phototubes.  The mirrors are
coated with aluminum and magnesium fluoride and have a measured
reflectivity of 90\% at 400~nm.  The mirror array is mounted with
individual adjustment on a plane of 3.0~cm Rohacell foam reinforced
with fibre glass epoxy films.  The average effective particle path
length in the gas viewed by the mirrors is 0.95~m.

The phototubes (Burle 8854) have 12.7~cm diameter photocathodes, and
are fitted with Hinterberger-Winston light cones to maximize the light
collection.  The funnels have an entrance diameter of 21.7~cm and can
accept light at angles of up to 26~degrees from the optical axis. They
are coated with aluminium to give a reflectivity of 85{\%} at 400~nm.
Magnetic shielding for the phototubes is provided by three concentric
$\mu$-metal shields and a soft iron housing. The photocathode faces are
coated with a layer of 200~$\mu$gm/cm$^2$ of p-Terphenyl and a
protective coating of magnesium fluoride. This wavelength
shifter increases the photoelectron yield \cite{CER-Baillon}.
The mean number of photoelectrons for a $\beta$=1 particle with a pure
nitrogen radiator was measured to be slightly less than 3.  LEDs
mounted on each Winston cone provide an on-line gain monitoring
system.  Calibration spectra are accumulated during normal running 
with the LED amplitudes set sufficiently low so that strong 
single-photoelectron peaks are resolved cleanly.  These peaks provide 
an accurate measure of the gain per photoelectron for each channel.

\begin{figure}
\hskip 0.1cm \psfig{figure=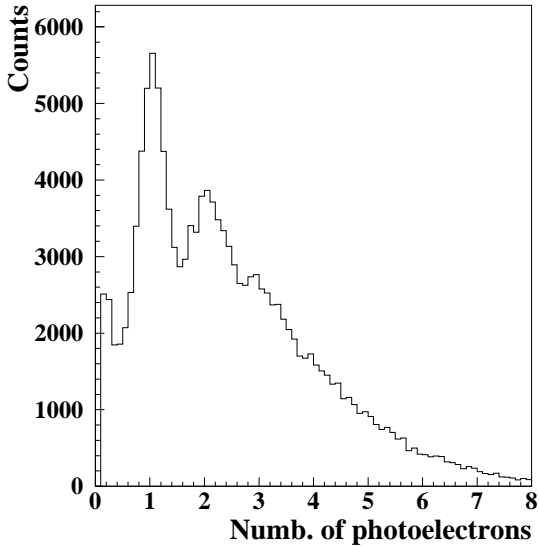,width=8.0cm}
\caption{Response of the \v{C}erenkov detector from 1995 data; note that
events below 0.15 photoelectrons have been suppressed.
\label{fig:cer}}
\end{figure}

The typical response of the \v{C}erenkov detector is shown in
Fig.\ref{fig:cer}.  The single photoelectron peak is prominent and
reasonably well separated from the pedestal.

\subsection{The Transition Radiation Detector}

The purpose of the transition radiation detector (TRD) is to provide a
pion rejection factor of at least 100 for 90\% positron efficiency at
5~GeV and above.

Only positrons produce transition radiation in the HERMES energy
regime and the detection of one or more TR X-rays in coincidence with
the charged particle can be used to discriminate between positrons and
hadrons.  The X-ray is difficult to distinguish because it is emitted
at an angle of 1/$\gamma$ to the momentum of the charged
particle and is therefore not separable in the detector from the
positron.

Because the space restrictions imposed on the TRD were not severe, a
relatively simple and economical design was chosen.  Rather than
optimize each module of the TRD for maximum performance, the overall 
performance is assured by the addition of extra modules.
The design of the TRD was optimized using a Monte Carlo
program that tracks charged particles and photons down to an energy
of 1~keV.  Care was taken to model properly the production of
$\delta$-rays in the radiator and the detector gas as well as hadronic
showers in the radiator.  Elements of the TRD simulation were later
integrated into the general spectrometer simulation
\cite{TRD-emerson}.

The HERMES TRD consists of 6 modules above and below the beam.  Each
module contains a radiator and a Xe/CH$_4$ filled proportional
chamber.  In addition, there are two flush gaps on either side of each
detector through which CO$_2$ is flowed to reduce the diffusion of
oxygen and nitrogen into the chamber.  Such contaminants would affect
the response significantly.  CO$_2$ is used in the flush gaps because
it is easily removed from the chamber gas during re-circulation.

The use of polyethylene foils as a radiator is not practical for
HERMES because of the large size of the modules (active area: 325 x 75
cm$^2$).  It is not possible to maintain a uniform separation of the
foils for such a large radiator.  A good approximation to a foil
radiator is provided by a pseudo-random but predominantly
two-dimensional matrix of fibers \cite{TRD-fibers}.  Several radiators
were compared in a test beam at CERN: foils, fiber matrices with
different fiber diameters, and polyethylene foams.  Fiber radiators
were shown to produce a response only slightly less than that from a
foil radiator.  The final design uses fibers of 17-20~$\mu$m diameter
in a material with a density of 0.059~g/cm$^3$.  This density is
significantly less than would be found to be optimal if hadronic
showers and delta rays from the radiator were ignored.  The radiators
are 6.35~cm thick, which corresponds to an average of 267 dielectric
layers.

The detectors are proportional wire chambers of conventional design,
each of which consists of 256 vertical wires of 75~$\mu$m gold coated
Be-Cu separated by 1.27~cm.  The wires are not soldered but rather held
in crimped pins.  This makes it much simpler to replace a broken wire
in situ.  The wires are positioned with 25~$\mu$m accuracy using
precisely machined holes for the crimp pins in plastic strips (Ultem)
that are glued to the aluminium chamber frames.  The chambers are
2.54~cm thick.  Xe/CH$_4$ (90:10) is used as the detector gas because
of its efficient X-ray absorption.  The electric field in the chamber
is produced by a +3100~V potential applied to the anode wires; the
cathode foils are at ground potential.  The large voltage on the anode
wires is needed to produce an electric field in the critical region
between the wires of sufficient strength to collect the ionization
electrons within the 1~$\mu$sec ADC gate used in the readout
electronics.  The wire diameter was chosen to be unusually large to
allow operation at this high voltage while limiting the gas gain to
about 10$^4$. The drift cell properties were optimized using the
computer program GARFIELD \cite{BC-garfield}.

The active gas volume of the detector is defined by the cathodes which
are 50~$\mu$m thick mylar foils aluminized on both sides.  The foils
are stretched to an unusually high tension (1.75~kN/m) to make them more
stable against pressure differences between the detector and the flush 
gaps.  This is important because a 10~$\mu$m deviation in the position
of the cathode causes a 1\% gain shift in the detector.  All frames
supporting the anode wires and the cathode foils were 
pre-stressed during construction.

The tension of each wire was measured using resonant vibration induced by
the combination of a current flowing through the wire and an external magnetic
field.  Furthermore, each wire was scanned with an X-ray source every
2.54~cm during operational tests after assembly.
Any gain variations of more than 10\% caused by imperfections in the wire
resulted in the wire being removed and re-strung.  

The signals from the wires are amplified by electronics mounted along
the long edge of the detector furthest from the beam.  Balanced
differential signals are transmitted over 35~m long twist/flat cables
to the electronics trailer.  The use of differential signals reduces
common mode noise, which would severely degrade TRD performance.
Receiver modules filter, amplify, and delay the signals by 256~ns
while a first level trigger decision is made, at which time they are
digitized by LeCroy 1881M Fastbus ADC's.  Test pulses are injected via
an antenna at the other end of the wires to allow for a quick check of
the detector and readout electronics without beam.

Since Xe is expensive, the chamber gas must be recirculated and
purified.  Impurities in the gas are kept at acceptable levels by
molecular sieves and activated copper (N$_2 \approx$~1\%, CO$_2
\approx$~0.25\%, O$_2 \approx$~ppm).  As a further precaution to keep
impurities low, all detectors were He leak checked during
construction.  This was particularly important to identify and plug
leaks in the crimped pins used to fix the wires.  A sophisticated
pressure control system based on an industrial programmed controller
adjusts the flows into the detector and the gaps in
such a way that the differential pressure between these volumes is
less than a few $\mu$bar \cite{TRD-thiessen}.  A pressure stability
of 1-2 $\mu$bar corresponds to a gain stability of a few percent.  The
chambers track atmospheric pressure in order to avoid using very thick
foils to contain the gas.  However, experience has shown that even
during periods of large atmospheric pressure variations, the
differential pressure between the detector volume and the flush gaps
is maintained below 1-2~$\mu$bar by the control system.  The gas
system is described in more detail in \cite{TRD-thiessen}.

\begin{figure}
\hskip 0.01cm \psfig{figure=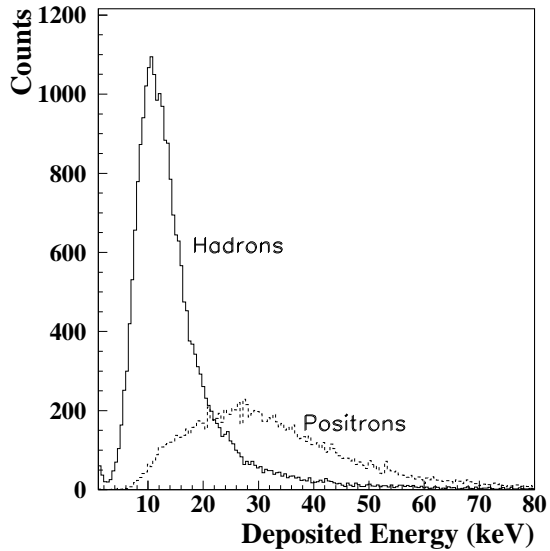,width=8.0cm}
\caption{Response of a single TRD module to positrons and hadrons,
integrated over all momenta.  Data from 1995 were used for this plot.
\label{fig:trdfig1}}
\end{figure}

Both positrons and hadrons deposit energy in the detector due to the
ionisation of the chamber gas ($dE/dx$).  The most probable energy
deposited by a 5~GeV pion is about 11~keV.  Positrons deposit on
average approximately twice this amount of energy due to transition
radiation and the relativistic rise in $dE/dx$. The response to
positrons and hadrons in one module is shown in Fig.~\ref{fig:trdfig1}.
Pure positron and hadron samples are selected by placing stringent,
inefficient conditions on the other PID detector responses.  The
distributions are very broad and it is clear that information from
several modules must be combined for high quality hadron rejection.
In particular the long, high energy tail in the hadron distribution
overlaps significantly with the positron distribution.
\begin{figure}
\vskip -0.25cm
\hskip 0.01cm \psfig{figure=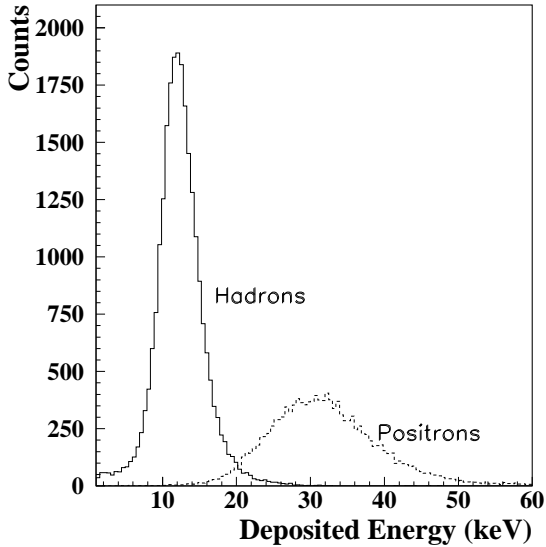,width=8.0cm}
\caption{Truncated mean for six modules for positrons and hadrons,
integrated over all momenta.  Data from 1995 were used for this plot.
\label{fig:trdfig2}}
\end{figure}
A simple but effective method of analysis is the truncated mean
method.  In this technique, the largest signal from the six modules is
discarded and the average of the remaining five modules is used.
This procedure reduces the mean of the positron distribution but has a
much more significant effect on the hadron tail that is due to rare
events - production of energetic knock-on electrons.
Fig.~\ref{fig:trdfig2} shows the truncated mean distributions for the
HERMES TRD derived from 1995 data.  It is clear that the separation of
the hadrons and positrons is greatly enhanced.  The pion rejection
factor varies with energy because both the energy loss by charged
particles and the production of transition radiation depend on the
energy of the incident particle.  It is customary when describing the
performance of a TRD to place a cut on the response of the detector
such that 90\% of the positrons are above the cut.  The cut is set for
higher efficiency for physics analysis.  Using the truncated mean
method, the energy-averaged PRF is about 150 for 90\%
positron efficiency.  If the data are analysed as a function of
momentum, a PRF of 130 is deduced at 5~GeV, exceeding the design goal.
The PRF is somewhat smaller for lower momenta (80 at 4~GeV) and
larger for higher energies (up to 150).

\begin{figure}
\vskip -0.25cm
\hskip 0.01cm \psfig{figure=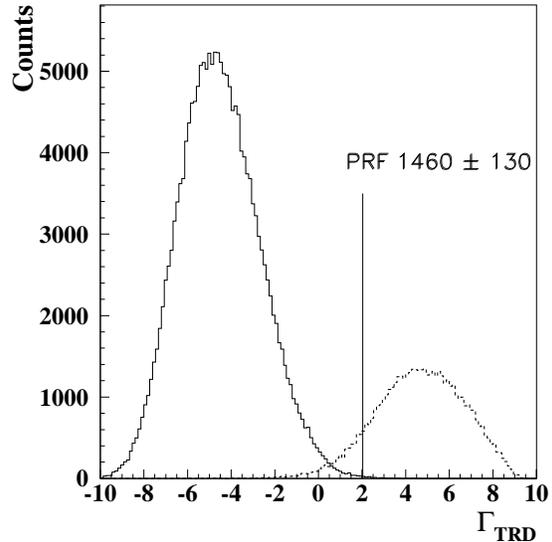,width=8.0cm}
\caption{The logarithmic likelyhood $\Gamma_{TRD}$, integrated
over all momenta, for hadrons (positrons) is plotted as the solid line 
(dashed line).  Data from 1996 were used for this plot.  The vertical line
indicates the location of a 90\% efficiency positron cut.
\label{fig:trdfig3}}
\end{figure}

The PRF can be improved using a probability based analysis similar to
the one described in the section on PID system performance below.  A
probability function can be defined as

\centerline{$\Gamma_{TRD}$= log$_{10}[P_{e^+}/P_h$],}

\noindent
where $P_{e^+}$ and $P_h$ are the probabilities that a particle was a positron
or a hadron respectively.  The result of such an analysis for 1996 data is 
shown in Fig.\ref{fig:trdfig3}.
A cut for 90\% positron efficiency results in a PRF of 1460~$\pm$~130. 
Even at 95\% positron efficiency, the PRF remains very good (489$\pm$25).

\subsection{PID System Performance}

The responses of the four PID detectors are combined to provide good
hadron rejection.  This has been done in several ways.  The simplest
technique is to consider the responses individually, determine a cut
that separates positrons and hadrons in each detector, and define a
positron/hadron as a particle identified as such in all detectors.  A
very clean sample of a given particle type can be produced in this
way but at the cost of efficiency.  Typically, this technique is used
only for detector studies.  The efficiency of the system can be
improved while retaining good particle type separation by using a
probability based analysis.  The responses of the calorimeter, the
pre-shower detector, and the \v{C}erenkov detector are combined to
produce the quantity PID3 defined as

\centerline{PID3= log$_{10}$ $[ (P_{Cal}^e P_{Pre}^e
P_{Cer}^e) / (P_{Cal}^h P_{Pre}^h P_{Cer}^h) ]$,}

\noindent
where $P_j^i$ is the probability that a particle of type $i$ produced
a given response in detector $j$.  The $P_j^i$'s are determined by
comparing the detector responses for each track to typical response
functions for each detector called parent distributions.  The latter
are either derived from the data or modelled using Monte Carlo
techniques.  The use of momentum dependent parent distributions is
required to account for the varying responses of the detectors.  PID2
is defined in a similar way to PID3 using only the calorimeter and the
pre-shower detector.  PID4 including the TRD is also defined but will
be discussed later.

\begin{figure}
\hskip 0.01cm \psfig{figure=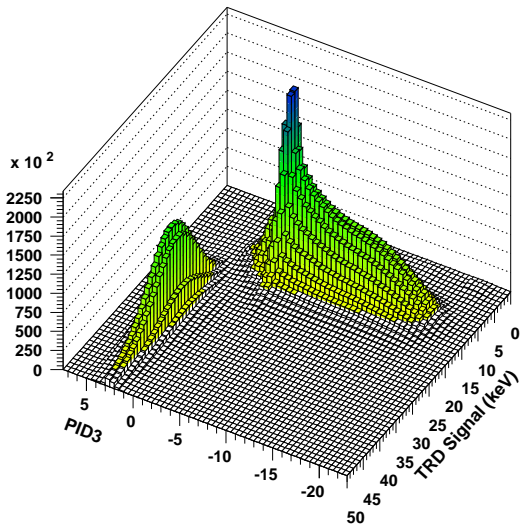,width=8.0cm}
\caption{Plot of the PID3-TRD plane.  Positrons appear at 
positive PID3 and large values of 'TRD Signal'.
\label{fig:pid3d}}
\end{figure}

For the 1995 data, PID3 was used in combination with the TRD truncated
mean to make a two-dimensional cut identifying positrons and hadrons.
This is shown in Fig. \ref{fig:pid3d} where it is clear that a clean
positron sample can be isolated at positive PID3 and large TRD
truncated mean (TRD Signal).  The advantage of using this technique
rather than combining all four detectors is that it allows detailed
systematic studies to be done on the probability analysis
\cite{PID-Kaiser}.  For example stringent cuts (low efficiency, but
high purity) can be applied to the TRD to define clean samples of
positrons and hadrons that can be used to study the PID3 response and
vice-versa.

\begin{figure}
\hskip 0.01cm \psfig{figure=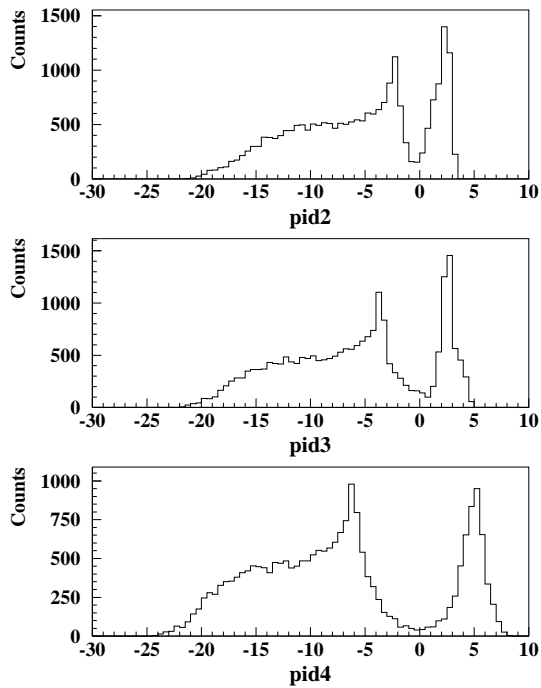,width=8.0cm}
\caption{Comparison between the PID parameters PID2, PID3 and PID4
for the 1995 data set.  Positrons have positive values of PID2, PID3,
and PID4, while hadrons have negative values for these quantities.
\label{fig:pid234}}
\end{figure}

One can incorporate the TRD into the probability
analysis to improve performance even further.  The resulting function
is labelled PID4.  However, one then loses the straightforward
systematic cross-checks described above and Monte Carlo simulation
must be used to estimate the contamination of each particle sample as
well as the detection efficiency.  A graphical comparison of PID2,
PID3, and PID4 is shown in Fig.~\ref{fig:pid234}.  In order to keep as
many events as possible for the final analysis, a so-called PID
downshifting scheme was implemented.  If all PID detectors provide valid
data, a cut on a 2-dimensional PID3-TRD plot is used, which is
equivalent to a straight cut on PID4.  However, if information from
either the TRD or the \v{C}erenkov detector is not available a more
limited scheme is used: if the TRD data are not available PID3 is
used, while if the \v{C}erenkov data are not useful a two-dimensional
cut on the PID2-TRD plane is considered.  For the analysis of the
inclusive data in 1995, the full PID3-TRD scheme was used for over 
97\% of the useful events.

The hadron contamination of the positron sample was determined in
several ways.  In the more limited PID schemes, a value can be derived
from clean particle samples obtained using the detector not considered
in that particular method.  However, for the full PID scheme, the
hadron contamination can only be estimated due to the fact that all
detectors are used in the analysis and there is no control detector to
define clean samples.  The different analysis techniques give
consistent results.  The most straightforward of these methods is a
simple fit to the total probability distributions using an assumed
shape for the tail for each particle sample in the overlap region, for
example a Gaussian or an exponential.  The values of the hadron
contamination for a given positron efficiency are listed in
Table~\ref{pidtable} for the $x$ bins used in the analysis of the 1995
data.  More details on the particle identification system can be found
in ref. \cite{PID-Kaiser}.

\begin{table}[h]\centering
\begin{tabular}{|l|l l|}\hline
$x$-Bin  & \multicolumn{2}{|c|}{PID3-TRD} \\ \hline
              & $e^+$ eff. & $h^+$ cont. \\ \hline 
0.023 - 0.04  & 97.77 & 1.18  \\  
0.04  - 0.055 & 98.38 & 0.81  \\  
0.055 - 0.075 & 98.78 & 0.55  \\
0.075 - 0.1   & 99.21 & 0.37  \\
0.1   - 0.14  & 99.44 & 0.22  \\ 
0.14  - 0.2   & 99.64 & 0.16  \\
0.2   - 0.3   & 99.71 & 0.16  \\
0.3   - 0.4   & 99.72 & 0.16  \\
0.4   - 0.6   & 99.72 & 0.11  \\
\hline
\end{tabular}
\caption{Positron Efficiency and Hadron Contamination in \% for the PID3-TRD
Method (1995).}
\label{pidtable}
\end{table}

\begin{figure}
\hskip 0.01cm \psfig{figure=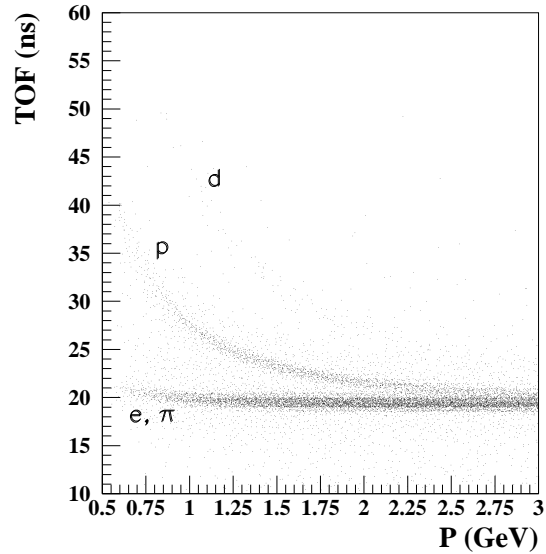,width=8.0cm}
\caption{Time of flight vs momentum for hodoscope H1.
\label{fig:pidtof}}
\end{figure}

Time of flight information from the hodoscopes can be used to separate pions,
kaons, protons, and deuterons at low momentum (up to $\approx$2~GeV).  The
time of flight is referenced to the accelerator bunch signal.  The
TDC values must be corrected for vertical position in the scintillators to
take into account the transmission time of the light.  The time resolution
is 300~ps.  Fig.~\ref{fig:pidtof} shows the time of flight as a function of 
momentum.  Protons, deuterons and perhaps some kaons are visible.

Another very effective way to do particle identification is to
reconstruct the mass of particles with decay products in the
spectrometer.  This has been done for $\pi^0$, $\rho$, $\eta$, K$_S$, $\phi$,
$\omega$, J/$\Psi$, and $\Lambda$.  More details on decaying particles
can be found in ref. \cite{PID-semisig}.

\section{The Luminosity Monitor}

The luminosity measurement is based on the elastic scattering of beam
positrons from target gas electrons $e^+e^- \rightarrow e^+e^-$
(Bhabha scattering) and the annihilation into photon pairs $e^+e^-
\rightarrow \gamma\gamma$.  The cross sections are known precisely,
including radiative corrections, e.g.\ $e^+e^- \rightarrow
e^+e^-\gamma$, $e^+e^- \rightarrow \gamma\gamma\gamma$ \cite{lumir2}.
In the future with an electron beam in HERA, the process $e^-e^-
\rightarrow e^-e^-$ (M{\o}ller scattering) will be used \cite{lumir4}.
The scattered particles exit the beam pipe at z~= 7.2~m and are
detected in coincidence by two small calorimeters with a horizontal
acceptance of 4.6 to 8.9 mrad, which is limited by the size of the
beam aperture in the magnet shielding plate.  For a beam energy of
27.5 GeV the symmetric scattering angle is 6.1 mrad and both scattered
particles have half of the beam energy. \\

\begin{table}[hbt]
\centering
\begin{tabular}[t]{|l|r|} \hline
Density $\rho$ (g/cm$^3$) & 7.57 \\ \hline
Radiation length $X_0$ (cm) & 1.03 \\ \hline
Moli\`ere radius $R_M$ (cm) & 2.38 \\ \hline
Critical energy $E_C$ (MeV) & 9.75 \\ \hline
Index of refraction $n$ & 2.15 \\ \hline
Optical transparency (nm) & $\ge 380$ \\ \hline
Radiation hardness (Gy) & $7 \cdot 10^5$ \\ \hline
\end{tabular}
\caption{Properties of NaBi(WO$_4$)$_2$ crystals. \label{lumit1}}
\end{table}

Due to the high radiation background in the region very near to the
beam, the calorimeter consists of \v{C}erenkov crystals of
NaBi(WO$_4$)$_2$ \cite{lumir6,lumir7,lumir8}, which have a very high
radiation hardness on the order of $7 \cdot 10^5$ Gy.  To further
minimize radiation damage, the calorimeters are moved away from the
beam pipe exit window by $\sim$ 20 cm in the horizontal direction for
beam injection and dumping.  The properties of NaBi(WO$_4$)$_2$ are
summarized in Table~\ref{lumit1}.  The small radiation length and the
small Moli\`ere radius allow a very compact calorimeter design.  Each
calorimeter consists of 12 crystals of size 22$\times$22$\times$200
mm$^3$ in a 3$\times$4 array.  The crystals are wrapped in aluminized
mylar foil and coupled with optical grease to 1.9~cm Hamamatsu R4125Q
photomultipliers with a radiation hard synthetic silica window and a
bialkali photocathode.  For monitoring the gain of the counters, each
crystal is fed with light pulses from an optical fibre of the HERMES
gain monitoring system described in section 6.2.  An energy resolution
of $\sigma$(E)/E $\simeq$ (9.3$\pm$0.1)\% / $\rm \sqrt(E)$ (E in GeV)
has been determined for a 3$\times$3 matrix from test beam measurements
with 1-6~GeV electrons. \\

\begin{figure}[htb]
\vspace*{-1cm}
\centerline{\psfig{figure=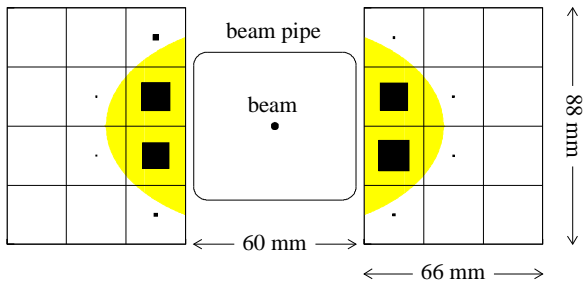,width=10cm}}
\noindent
\vspace*{-1cm}
\caption{Hit distribution in the calorimeters of the luminosity monitor.
         The size of the boxes is proportional to the number of hits per
         channel. The shaded area shows the beam pipe acceptance.
         \label{lumif1}}
\end{figure}

\begin{figure}[htb]
\vspace*{-1.5cm}
\psfig{figure=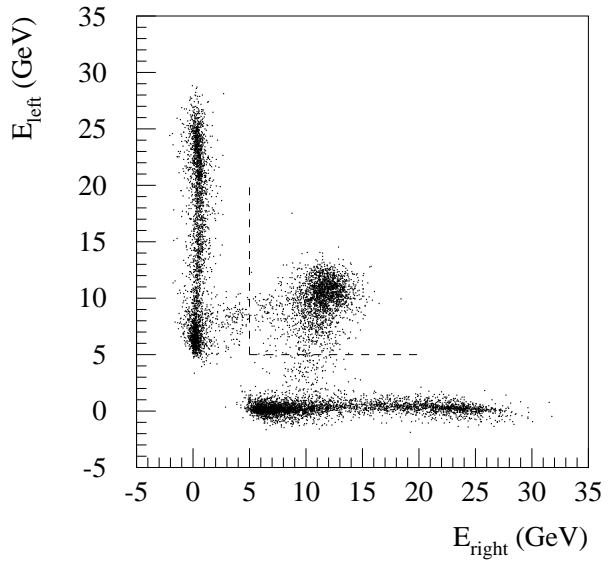,width=10cm}
\noindent
\vspace*{-0.5cm}
\caption{Scatter plot of the deposited energy in the left luminosity 
         detector versus the deposited energy in the right detector. 
         The dashed line shows the trigger threshold of 5 GeV in both
         detectors for Bhabha events.
         \label{lumif2}}
\end{figure}

Fig.~\ref{lumif1} shows the hit distribution in the calorimeters.
Most of the scattered particles hit the inner crystals near the beam
pipe.  The deposited energy is reduced by lateral shower leakage,
especially for hits near the inner crystal edge.  A scatter plot of
the energy in the left detector versus the energy in the right
detector is shown in Fig.~\ref{lumif2}.  Most of the background
events have a high energy deposition in only one of the detectors
while Bhabha events have a high energy deposition in both detectors.
They are separated from background by triggering on a coincident
signal with energy above 5~GeV in both the left and right calorimeter.
A Bhabha coincidence rate of 132~Hz was measured with this trigger
scheme for a beam current of 20~mA and a $^3$He areal target density of 
$1\cdot10^{15}$~nucleons/cm$^2$.  This provides a statistical accuracy
of the luminosity measurement of 1\%, within about 100~s. \\

\section{The Trigger}

\subsection
{Trigger Requirements}
     
The function of the trigger system is to distinguish interesting
events from background with high efficiency, and initiate digitization
and readout of the detector signals. HERMES requires physics triggers
corresponding to deep-inelastic positron scattering, photoproduction
processes (where no positron is detected) and additional triggers for
detector monitoring and calibration. The trigger hierarchy is
potentially capable of four levels.  However, for running in 1995-97
only the first-level trigger was required.

The first-level trigger decision is made within about 400~ns of the
event using prompt signals of the scintillator hodoscopes, the
calorimeter, and a few wire chambers. This can be divided into the
time necessary for signal formation in the detectors and
transportation to the electronics trailer ($\approx$~250~ns including
the transit time of the particle from the target), and the time needed
for a decision by the trigger electronics ($\approx$~150~ns).  The
first-level trigger initiates digitization by the readout
electronics. The second-level trigger could use any information
available within the 50~$\mu$s during which time a hardware clear of
the Fastbus readout electronics is possible. The third-level decision
could be made on a timescale of a few hundred micro-seconds by
existing digital signal processors in Fastbus, which would first read
out only that part of the detector information required for the
decision.  In fourth level, the full data stream could be filtered by a
farm of Intel Pentium Pro processors that can make a decision on a
timescale of $\leq$1~ms.  Only the first-level trigger is described
below.

The DIS trigger selects positron events by
requiring hits in the three scintillator hodoscopes, H0, H1, and H2,
(only H1 and H2 in 1995) together with sufficient energy deposited in
two adjacent columns of the calorimeter, in coincidence with the accelerator
bunch signal (HERA Clock).  The forward H0 scintillator was installed
in 1996 because backward-going particles from the HERA proton beam are
not adequately suppressed by the H1 -- H2 timing alone. With H0
included, the primary sources of background are hadrons either
produced from positron beam interactions in the upstream collimators,
or photoproduced in the target. The requirement of hits in the H0 and
H1 hodoscopes suppresses neutral particle background. The primary
charged background discrimination is done using a calorimeter
threshold set above the minimum ionizing energy deposition of
0.8~GeV. The calorimeter has high efficiency for electromagnetic
showers, but relatively low efficiency for hadronic showers. The
hodoscope thresholds were set below minimum ionizing. A calorimeter
threshold setting of 3.5~GeV (used for the 1995 running and
corresponding to $y\leq 0.85$) suppresses the charged hadronic
background rate by a factor of $\geq$ 10 in the trigger.  The typical
trigger rate at a luminosity of
$5~\times~{10}^{32}$~nucleons/cm$^2$/sec, was about 50~Hz in 1995.
The threshold was lowered to 1.4~GeV during 1996 to increase the $y$
range to $y\leq 0.95$ and improve the acceptance for semi-inclusive
particles.

The photoproduction trigger detects hadrons such as K, $\rho$, D$^0$,
J/$\psi$ and $\Lambda^0$ that are produced at low $Q^2$ and decay to
two or more charged particles.  Typically the scattered positron angle
is too small for detection. The trigger requires charged particle
tracks in both the top and bottom detectors, as identified by the
three hodoscopes and the BC1 chamber as well as the HERA Clock. The
back chamber requirement eliminates those showers originating in the
upstream collimators, which are confined near the beampipe and hit the
tips of the hodoscopes but not the wire chambers which are well
shielded by the magnet steel.

\subsection
{First-level Trigger Architecture}
     
The signals needed for the first-level triggers are the hodoscopes
(H0, H1, H2), a multiplicity of two or more hits in the back wire
chambers (BC1), and a cluster of calorimeter blocks with total energy
above threshold.  Hits in the top AND bottom halves of the spectrometer
are required for the photoproduction trigger.

Each hodoscope photomultiplier (PMT) signal is passively split with
one output going to a LeCroy 1881M analog to digital convertor (ADC)
and the other going to a LeCroy 3420 Constant Fraction Disciminator
(CFD).  The individual CFD outputs are fed to LeCroy 1875A time to
digital convertors (TDC). The high impedance OR outputs of the CFD's
for each hodoscope wall are connected together in a chain to form the
hodoscope logic signals.

Each calorimeter wall consists of 42 columns of 10 lead-glass blocks
each. PMT signals from each block in a given 10-block column are split to
provide ADC inputs, and added linearly to form a Column Sum using
custom-built NIM modules. Segment Sums of 20 blocks are formed by
linear addition of pairs of adjacent Column Sums, in LeCroy 428F
modules. At least one Segment Sum contains more than 95\% of the
energy of each electromagnetic shower.  All combinations of adjacent
columns are used to fully cover a calorimeter wall. The analog Segment
Sum signals are discriminated by CFD's to form TDC and logic signals.

The back chamber signals are used only for the photoproduction
trigger.  ECL signals are utilized from each of the 256 wires of the
top and bottom X-planes of BC1. LeCroy 4564 modules are used to
produce logical OR's of adjacent wires in groups of 16. The
multiplicity of 16-wire groups that fired is determined in a LeCroy
4532 module.

The resulting ECL logic signals are reshaped by a LeCroy 4415A
differential discriminator and collected into a signal bus. Relative
delay adjustments can be made with a LeCroy 4518 programmable CAMAC
logic delay module. The trigger logic is established in LeCroy 4508
programmable logic units (PLU), which allow changes to be made
easily. The signal bus provides the inputs to the PLU, which is strobed
with the HERA clock. The PLU ECL outputs are converted to NIM,
prescaled appropriately with CERN prescalers, and brought together in
the Master Event OR. The Master Event is retimed by the HERA Clock and
gated by the data acquisition NOT BUSY signal to produce the Master
Trigger that initiates digitization and readout of the event.

\subsection{Trigger Performance}

The purity of the deep inelastic scattering (DIS) trigger is good for
a high calorimeter threshold (3.5~GeV) and acceptable for a low
threshold (1.4~GeV).  For a 3.5~GeV calorimeter threshold, two
thirds of the triggers had tracks, 95\% of reconstructed tracks came
from the target, and one third had accompanying positrons.  For a
1.4~GeV calorimeter threshold, the physics trigger rate 
increased by a factor of about 6 while the fraction of DIS positrons
only went up by about 10\%.  These runs had significant
contamination coming from the collimator just upstream of the target.
About two thirds of the events had tracks, but only 70\% of the tracks
came from the target, with most of the remaining coming from the
collimator.

Approximately one third of the photoproduction triggers have
reconstructable tracks.  Of these, over 3/4 have two or more tracks,
due to the requirement of hits in the lower and upper part of the
detector simultaneously.  Only 6-7\% have positrons, which is not
surprising given that the aim of the trigger is to detect events at
low $Q^2$ where the positron goes down the beampipe.  However, most of
the tracks come from the target region.

During normal polarised running with the high calorimeter threshold,
15\% of the photo-production triggers were also DIS triggers.  Roughly
2.5 photo-production triggers were collected for each DIS trigger.
During polarised running with the low calorimeter threshold the overlap
increased to 40\% due in part to the increased rate in the high-y 
DIS triggers.  The number of photo-production triggers remained the same,
since the calorimeter is not in this trigger.  

\section{Data Acquisition and Readout Electronics}

The backbone of the data aquisition system is constructed in Fastbus.
It consists of 10 front-end crates, the event collector crate, and the
event receiver crate connected to the online workstation cluster via
2 SCSI interfaces. CERN Host Interfaces (CHI) act as Fastbus masters.
To enhance their readout performance they are equipped in most places
with Struck Fastbus Readout Engines (FRE), featuring one or two
Motorola 96002 DSPs.  The event collector in the electronics trailer
is connected to the event receiver crate in a computer room via a
Fibre Optical Link \mbox{(STR330/FOL)} plugged into the respective
CHIs.  Pairs of Fastbus crates, read via only one CHI/FRE, are formed
by cluster interconnects (CI), where overall readout timing considerations
allow for it.  The drift chamber signals are digitized using LeCroy
multi-hit multi-event 16 bit 1877 sparsifying multiblock 96 channel
TDCs with a resolution of \mbox{0.5 ns/channel}. Charge from the
various photomultiplier tubes and the TRD is digitized by LeCroy
multi-event 64 channel 1881M multiblock ADCs.  The resolution is
\mbox{50 fC/channel} and the full scale range is 13 bits above
pedestal.  During a multiblock transfer from 1881M ADCs or 1877 TDCs,
the CHI/FRE combination results in a readout time of \mbox{127
ns/word} (CHI without FRE: \mbox{270 ns/word)}, where the slave timing
is contributing \mbox{90 ns/word} and the FRE \mbox{37 ns/word}.  The
multiblock passing time (time for two neighbouring modules to pass the
token) is found to be 640 ns. Reading beyond a cluster interconnect
adds a penalty of about \mbox{50 ns/word}. This results in a block
read speed of more than 30 MB/s in crates having a master and more
than 20 MB/s in crates beyond a CI. As soon as first data are
collected into the FRE's input FIFO, the active DSP starts processing
the incoming data, writing the output to its output FIFO. The LeCroy
multi-event 10 bit 64 channel 1875A was choosen for digitizing the
time of flight signals. The resolution is set to 25 or \mbox{50
ps/channel} depending on the detector. As only six 1875A's are read out
in total, the lack of multiblock capability on this unit is only a
minor disadvantage.  The magnet chamber readout is instrumented with
the LeCroy VME based PCOS4 system, consisting of one 2749 and twelve
2748 modules. The readout time (including the transfer to the event
collector memory) is \mbox{450 ns/hit wire}. Vertex chamber data
arrive from the detector as a 16 bit ECL \mbox{STR330/ECL} data
stream, one 16 bit word containing the plane and strip number of one
hit strip. Incoming data are passed in direct mode from the Struck ECL
interface to the associated DSPs, where hot channels are suppressed
using a linear 16 bit lookup table.  The trigger pattern is available
to all front ends, to allow for selective readout.  Event collection
is done from each FRE output FIFO via a cable segment, connecting
directly to the FRE cable port. Double buffering is implemented in all
essential places to provide a low front-end deadtime. Event collection
is done in parallel with the next event readout with incoming events
processed alternately between two DSPs in each of the FREs. A typical
overall reduction in event size from $\approx$18 to \mbox{10 KB/event} is
achieved at the DSP level. 

In addition to the standard detector readout, there is implemented a
variety of asynchronous independent events, capable of trigger rates
exceeding 5~kHz, including the luminosity monitor and various high
speed calibration and monitoring equipment. Depending on the
application, the data from these events are either collected by the
event collector during idle time or by dedicated additional CHIs and
FREs. They can be defined either as so-called user events, or as a set
of histograms read directly from CHI RAM or FRE DSP BRAM where they
have been accumulated.

One VME branch with 4 crates and three CAMAC branches with a total of
9 crates are connected to the event builder crate to handle special
data acquisition tasks, such as obtaining the electron bunch number, and
slow control.  The VME interface is a CHI/VSB2 module and VDB. The
VME crates are connected via a STR723 in combination with a
STR725. The CAMAC crates are controlled by standard CCA2 type crate
controllers.  Scaler information is obtained from STR200 32 channel 32
bit Fastbus scalers.

The code for both the 68030 and the 96002 is written in assembly
language. An online library of Standard Fastbus, CAMAC and VME
procedures for jobs that are not time-critical can be called from
FORTRAN or C in the workstation cluster. These procedures are executed
in the respective front-end CHIs.  All event data and the slow control
information are available for online monitoring and analysis in the
online data stream, and are easily available to code in high level
languages.

Data are written to 9 GB staging disks over the course of a fill of
the storage rings, typically lasting 8-12 hours. The data are copied
between fills to storage silos on the DESY main site via a FDDI link
and associated hardware.  In parallel, they are stored on local
Exabyte tape drives for redundancy.  The local system was switched to
DLT tapes late in 1996.  A total of 2~TB was recorded in 1995 and
4.1~TB in 1996.  The deadtime during standard running is typically
well below 10\% and the downtime due to the data aquisition system is
estimated to be below 1\%.  The possible throughput of \mbox{1.5 MB/s}
was determined by the CPU and I/O bandwidth of the event distributing
work station (Alpha 3000X), which corresponds to a 150 Hz event rate
at the average event size.  By replacing the 175 MHz 3000X with a 266
MHz 5/266, this was doubled for 1997.

\section{Beam Diagnostics and Tuning}

Since HERMES is a fixed target experiment, the scattered particles are
mostly forward and the detectors are necessarily in close proximity to
the beamline.  This makes the experiment sensitive to beam
induced background.  The two main sources of background associated
with the positron beam are synchrotron radiation and showers from halo
particles striking the collimators just upstream of the target.
Several closed orbit corrections (i.e. that affect the beam only
locally) are available to help minimize background.  These are of two
types: symmetric which affect the position of the beam at the
interaction point (IP), and asymmetric which vary the angle at the IP.
It was determined to be crucial to tune the beam in the straight
section upstream of the experiment to minimize synchrotron radiation.
In particular, background from this source is acceptable only when a
special asymmetric orbit correction well upstream of the IP is
optimized.  The detectors most susceptable to backgrounds are the
front tracking chambers and the TRD, the former because they must
operate in front of the magnet and hence in the presence of a large
flux of low energy particles, the latter because it is particularly
sensitive to photons by design.  The TRD was made less vulnerable to
synchrotron radiation by adding a thin sheet of lead ($\ll$1~mm) in
front of the first chamber to absorb some of the low energy incident
photon flux.

To help optimize background conditions before the detectors are turned
on, dedicated scintillators were installed around the beam pipe.  This
system is called the Tuning Scintillation Telescope (TST).
The TST consists of 2 sets of scintillators (8 each) that are located
upstream of the front chambers at z~= +1.4~m and behind the
luminosity detector at z~= +8.3~m. Each set has two layers of 4
scintillators surrounding the beampipe to view the beam halo. The
first layer is unshielded and therefore is sensitive to synchrotron
radiation as well as charged particles. The second layer is shielded
with 3 mm of lead, so the synchrotron radiation is mostly suppressed.
Charged particle halo and synchrotron radiation can be identified
using coincidences and anticoincidences between the first and second
layers of scintillators.  Beam tuning is then a matter of
minimizing the counting rates and making them symmetric.  The TST is
used primarily during initial beam tuning when the accelerators are
first turned on at the start of a running period, and at the beginning
of a fill of the storage rings.

Beam position monitors (BPM) on either side of the IP allow the beam
to be adjusted to a pre-determined position and angle in a short time.
The BPMs are particularly useful at the beginning of a fill of the
storage rings.

The experiment is also exposed to particles that come from showers
originating from the proton beam.  A set of scintillators was
installed around the proton beam pipe just behind the calorimeter
to monitor this background.  Background from this source varies
significantly from fill to fill and even during a fill.  However,
essentially none of it is correlated with the events of interest so
its main effect is to increase the trigger rate and decrease the
proportion of good events going to tape.  As outlined previously, the
addition of the front trigger hodoscope (H0) has significantly
improved the situation.

\section{Summary}

This paper describes the spectrometer built to measure the products of
deep inelastic scattering of polarised positrons from polarised
internal gas targets at HERA.  After a relatively short commissioning
phase, the experiment started recording data in 1995.  All essential
detectors have performed well with only a few component failures over
the first two and a half years of running.  The acceptance of the
spectrometer allows the detection of hadrons in coincidence with the
scattered positron.  These important semi-inclusive measurements with
the identification of pions and kaons are unique to HERMES and will
allow the determination of the spin dependent quark distribution
functions.

\section{Acknowledgements}

The HERMES collaboration would like to thank all those who participated
in the design and construction of the spectrometer.  We are
particularly indebted to the MEA staff at DESY, many more of whom 
contributed than are mentioned in the author list.  We also gratefully
acknowledge the DESY management for its support. 

This project has been supported by 
the FWO-Flanders, Belgium;
the Natural Sciences and Engineering Council of Canada;
the Dutch Foundation for Fundamenteel Onderzoek der Materie (FOM);
the German Bundesministerium f\"ur Bildung, Wissenschaft, Forschung
und Technologie; the German Academic Exchange Service (DAAD);
the Italian Istituto Nazionale di Fisica Nucleare (INFN);
Monbusho International Scientific Research Program, JSPS, and Toray
Science Foundation of Japan;
the United Kingdom Particle Physics and Astronomy Research Council;
the U.S. Department of Energy and National Science Foundation;
and INTAS, HCM, and TMR contributions from the European Community.

\vskip 0.25cm

\noindent
* The theses mentioned in these references are available as HERMES 
internal notes, and can be accessed on the HERMES web pages:  \\
http://www-hermes.desy.de/notes/

\ \

\noindent
$^\dag$Deceased.


\begin{thebibliography}{999}
\bibitem{VC-D263} D-263 glass is produced by Desag, Germany.
\bibitem{CAL-f101} The F101 glass blocks were provided by the Lytcarino
factory, 140061 Lytcarino, Russia.
\end{thebibliography}
\end{document}